\documentclass[aps,a4paper, preprint, superscriptaddress,preprintnumbers,floatfix,nofootinbib,amsmath,amssymb,table]{revtex4-1}
\usepackage{hyperref}
\usepackage{cleveref}
\usepackage{subfig}

\usepackage{amstext,amssymb}
\usepackage[section]{placeins}
\usepackage{amsmath}
\usepackage{graphicx}
\usepackage{xspace}
\usepackage{color}
\usepackage{float}
\usepackage{comment}
\usepackage{amstext,amssymb}
\usepackage{amsmath}
\usepackage[section]{placeins}
\usepackage{xspace}
\usepackage{units}
\usepackage{array}
\usepackage{amsmath,bm}
\usepackage{amssymb}
\usepackage{gensymb}
\usepackage{soul}
\usepackage{textcomp}
\usepackage{slashed} 
\usepackage{multirow}
\usepackage{hyperref}
\usepackage{mathrsfs}
\usepackage{enumitem}
\usepackage{graphicx}
\usepackage{comment}
\usepackage{epstopdf}
\usepackage{float}
\usepackage{units}
\usepackage{feynmp}
\usepackage{slashed}
\usepackage{amsmath,bm}
\usepackage[table]{xcolor}
\usepackage[normalem]{ulem}
\definecolor{antiquewhite}{rgb}{0.98, 0.92, 0.84}
\definecolor{aqua}{rgb}{0.0, 1.0, 1.0}
\definecolor{amethyst}{rgb}{0.6, 0.4, 0.8}
\definecolor{applegreen}{rgb}{0.55, 0.71, 0.0}
\definecolor{byzantine}{rgb}{0.74, 0.2, 0.64}
\definecolor{cadetgrey}{rgb}{0.57, 0.64, 0.69}
\definecolor{candypink}{rgb}{0.89, 0.44, 0.48}
\usepackage{lipsum}
\usepackage{cals}
\usepackage{slashed} 
\newcommand{\be}{\begin{equation}}
\newcommand{\ee}{\end{equation}}
\newcommand{\bea}{\begin{eqnarray}}
\newcommand{\eea}{\end{eqnarray}}
\newcommand{\nn}{\nonumber}

\def\s1{\hat s}

\makeatletter

    \def\CT@@do@color{%
      \global\let\CT@do@color\relax
            \@tempdima\wd\z@
            \advance\@tempdima\@tempdimb
            \advance\@tempdima\@tempdimc
    \advance\@tempdimb\tabcolsep
    \advance\@tempdimc\tabcolsep
    \advance\@tempdima2\tabcolsep
            \kern-\@tempdimb
            \leaders\vrule
                    \hskip\@tempdima\@plus  1fill
            \kern-\@tempdimc
            \hskip-\wd\z@ \@plus -1fill }
    \makeatother

\usepackage{slashed}
\definecolor{ashgrey}{rgb}{0.7, 0.75, 0.71}
\definecolor{aureolin}{rgb}{0.99, 0.93, 0.0}
\definecolor{babypink}{rgb}{0.96, 0.76, 0.76}
\definecolor{buff}{rgb}{0.94, 0.86, 0.51}
\definecolor{chamoisee}{rgb}{0.63, 0.47, 0.35}
\definecolor{chartreuse(web)}{rgb}{0.5, 1.0, 0.0}
\definecolor{citrine}{rgb}{0.89, 0.82, 0.04}
\definecolor{emerald}{rgb}{0.31, 0.78, 0.47}
\definecolor{fawn}{rgb}{0.9, 0.67, 0.44}
\definecolor{fulvous}{rgb}{0.86, 0.52, 0.0}

\newcommand{\nua}[1]{\ensuremath{\rlap{\kern-2.5pt\ensuremath{\overset{\scriptscriptstyle(-)}{\phantom{\nu}}}}{\ensuremath{{\nu}_{#1}}}}\xspace}

\begin{document}
\title{Determination of neutrino mass ordering from Supernova neutrinos with T2HK and DUNE}
\author{Papia Panda}
\email{ppapia93@gmail.com}
\affiliation{School of Physics,  University of Hyderabad, Hyderabad - 500046,  India}
\author{Monojit Ghosh}
\email{mghosh@irb.hr}
\affiliation{Center of Excellence for Advanced Materials and Sensing Devices,
Ruder Bošković Institute, 10000 Zagreb, Croatia}
\author{Rukmani Mohanta}
\email{rmsp@uohyd.ac.in}
\affiliation{School of Physics,  University of Hyderabad, Hyderabad - 500046,  India}

\begin{abstract}

In this paper, we study the possibility of determining the neutrino mass ordering from the future supernova neutrino events at the DUNE and T2HK detectors. We estimate the expected number of neutrino event rates from a future supernova explosion assuming Garching flux model corresponding to different processes that are responsible for detecting the supernova neutrinos at these detectors. We present our results in the form of $\chi^2$, as a function of supernova distance.  For a systematic uncertainty of 5\% in normalisation as well as energy calibration error, our results show that, the neutrino mass ordering can be determined at $5 ~\sigma$ C.L. if the supernova explosion occurs at a distance of 42.7 kpc for T2HK and at a distance of 15.2 kpc for DUNE. Our results also show that the sensitivity of DUNE and T2HK get affected by the systematic uncertainties for the smaller supernova distances. Further, we show that in both DUNE and T2HK, the sensitivity gets deteriorated to some extent due to presence of energy smearing of the neutrino events. This occurs because of the reconstruction of the neutrino energy from the energy-momentum measurement of the outgoing leptons at the detector. 

\end{abstract}
\maketitle
\flushbottom

\section{Introduction }
\label{introduction}

When a massive star having mass greater than 8 $M_\odot$ with $M_\odot$ being mass of the Sun, comes to the end of its life, often the core of the star collapses and it explodes with huge energy and luminosity. This is called core-collapse supernova \cite{Burrows:2000mk,Horiuchi:2017qja}. In a core-collapse supernova, neutrinos of different flavours are produced in the energy range around few tens of MeV \cite{Bethe:1990mw}. In the first tens-of-milliseconds of the collapse, $\nu_e$ is produced from the electron capture. This phase of the supernova is called neutronization or the breakout burst. After this during the accretion phase, which lasts tens to hundred of milliseconds long, production of the electron flavour  neutrinos dominates. Finally in the cooling phase, neutrino-antineutrino pair of all three flavours are produced. The cooling phase lasts for a few tens of seconds \cite{Scholberg:2012id}. Neutrinos coming out from such supernova take $99 \%$ of its total gravitational energy after the explosion, while optical photons take only $ \sim 1 \%$. It is very interesting to note that neutrinos produced in the supernova reach earth before the optical photons \cite{Giunti:2007ry}. Therefore, supernova neutrinos can be an early signal for astrophysicists hinting a supernova burst. Supernova neutrinos can be a perfect information mediator for a black hole or a neutron star as a remnant of the massive star. With the help of the timing of supernova neutrino, one can also study gravitational wave (GW)~\cite{Deng:2023seh}. It has been shown that, without the information on the timing of supernova neutrino, the value of signal to noise ratio (SNR) for GW is $\sim 3.5$ whereas its value becomes 7 with neutrino timing~\cite{halzen2009reconstructing, Halzen:2009sm}. In this paper, we will study the oscillations of the supernova neutrinos. In the supernova there can be flavour conversion of the neutrinos when they propagate through the dense medium of supernova. This phenomenon gives us the opportunity to measure certain properties associated with neutrino oscillation. In the last millennium, there are several supernovae exploded in our Milky Way galaxy and in Large Magellanic Cloud. However, till now we are only able to detect neutrinos from SN1987A supernova, which occurred in Large Magellanic Cloud at a distance of 50 kpc from Earth \cite{DedinNeto:2023hhp,Lagage:1987xu}. Three detectors on earth detected the neutrino signal coming from SN1987A supernova almost $2.5$ hours prior to the optical detection. These detectors are: the water Cherenkov detector Kamiokande II \cite{Kamiokande-II:1987idp}, the water Cherenkov detector Irvine-Michigan-Brookhaven (IMB)~\cite{IMB:1987klg} and the scintillator detector Baskan~\cite{Novoseltseva:2009cr}. Although, in combination of three detectors, only a total 24 neutrino events were detected, they provided very important information related to astrophysics and physics in general. Due to the increased sensitivity of the present generation experiments, if a supernova burst occurs in near future, we will be able to detect a large number of neutrino events. This will provide an excellent opportunity to understand various phenomena related to neutrino physics.

In the phenomenon of neutrino oscillation, neutrinos change their flavors as they propagate through space and time. An immediate consequence of neutrino oscillation is that, neutrinos do have small but nonzero mass, with  three mass eigenstates $\nu_1, \nu_2$ and $\nu_3$ having masses $m_1$, $m_2$ and $m_3$, respectively. Quantum mechanical calculation of neutrino oscillation provides six oscillation parameters: three mixing angles ($\theta_{12}, \theta_{13}$ and $\theta_{23}$), two mass-squared differences: $\Delta m_{21}^2 = m_2^2 - m_1^2$, $\Delta m_{31}^2 = m_3^2 - m_1^2$ and one CP violating phase $\delta_{\rm CP}$ \cite{Esteban:2020cvm}. Among these six parameters, only three of them ($\theta_{12}, \theta_{13}$ and $\Delta m_{21}^2$) are measured very precisely leaving three oscillation parameters unknown, i.e., (i) value of $\theta_{23}$, (ii) sign of $\Delta m_{31}^2$ and (iii) allowed range of $\delta_{\rm CP}$. There are many current and future experiments which aim to measure these unknowns very precisely. In this work, we aim to estimate the sensitivity of the two future detectors, the DUNE detector in USA \cite{DUNE:2020mra} and the Hyper-K (HK) detector in Japan \cite{Hyper-Kamiokande:2016srs}, to determine the correct mass ordering of the neutrinos from the analysis of the supernova neutrinos. Currently, two possible orderings for the neutrinos are allowed, i.e., a mass spectrum with positive $\Delta m_{31}^2$, known as normal ordering and a mass spectrum with negative $\Delta m_{31}^2$, referred to as inverted ordering. The main purpose of DUNE and HK is to study the phenomenology of neutrinos produced from  accelerators. However, these experiments also provide an opportunity to study neutrinos from a future supernova burst. Because of the different detector technologies in DUNE and T2HK, the supernova neutrino detection methods will be different in these two experiments. In our study, we explore these aspects in detail. 

For the calculation of the fluxes of the supernova neutrinos, there are several models based on different assumptions~\cite{Totani:1997vj,Gava:2009pj,Burrows:2020qrp,Kuroda:2020pta,Nakazato:2012qf,DAmico:2019hih,Sukhbold:2015wba,Tamborra:2014hga,Walk:2019miz,Warren:2019lgb,Zha:2021fbi, DUNE:2020zfm}. Among all of these, Livermore \cite{Totani:1997vj} is the first ever supernova model where 1-D numerical simulation has been performed based on SN1987A supernova neutrino data. It gives the flux of neutrinos from onset of collapse upto 18 seconds after the core bounce. Another supernova model is GKVM (Gava-Kneller-Volpe-McLaughlin) \cite{Gava:2009pj}, which uses collective effects, shock wave effects for neutrino propagation in its calculation for the first time. GKVM model uses $S$ matrix formalism as well as hydro dynamical density profiles for supernova neutrino simulation. Another more realistic supernova model is Garching electron-capture supernova (ECSN) model \cite{Hudepohl:2009tyy}. In this model, a $8.8 M_\odot$ electron-capture supernova is simulated in spherical symmetry framework. This framework has been used throughout the supernova evolution to complete deleptonization of the forming neutron star. In the present work, we will do our analysis within the framework of Garching model and we will present the neutrino mass ordering sensitivity ($\chi^2$) as a function of supernova distance.

Study of supernova neutrinos in the context of both the mass orderings has been done in the past \cite{Dasgupta:2008cd,Wu:2014kaa,Scholberg:2017czd,Linzer:2019swe,Dasgupta:2008my,Dasgupta:2009mg,Gil-Botella:2016sfi,Das:2017iuj,Hyper-Kamiokande:2021frf,DedinNeto:2023hhp,Dighe:2003be,Dighe:1999bi,Chiu:2013dya,Brdar:2022vfr,Serpico:2011ir,Choubey:2010up}. In Ref.~\cite{Wu:2014kaa}, event rates corresponding to supernova neutrinos for a DUNE-type detector and HK-type detector are estimated using a spherically-symmetric supernova model with an 18 $M_\odot$ progenitor. Ref.~\cite{Scholberg:2017czd} shows the capability of current and future neutrino experiments to determine neutrino mass ordering in terms of event rates considering the different phases of the supernova burst. Ref.~\cite{Linzer:2019swe} measures the direction of the supernova neutrinos by triangulation method \cite{Brdar:2018zds}  i.e., relative neutrino arrival times at different detectors around the globe, for different neutrino mass ordering assumption. Using a toy model, Ref.~\cite{Das:2017iuj} shows the effect of non-standard self interactions (NSSI) on mass hierarchy determination in core-collapse supernova. The HK collaboration also calculated number of expected events for both the mass orderings assuming five different supernova neutrino flux models \cite{Hyper-Kamiokande:2021frf}. For the event rates corresponding to normal and inverted orderings by DUNE collaboration using the ECSN flux model, see Ref.~\cite{DUNE:2020zfm}. The possibility of measuring neutrino mass ordering at IceCube including the Earth matter effects of the supernova neutrinos in the context of Livermore and Garching models is discussed in Ref.~\cite{Dighe:2003be}. However, to the best of our knowledge,  this work would quantify for the  first time, the neutrino mass ordering sensitivity of supernova neutrinos by performing a detailed $\chi^2$ analysis in the context of Garching model for DUNE and HK detectors. In this regard, we will study (i) the effect of different detection methods of the supernova neutrinos in DUNE and HK detectors, (ii)  effect of systematic uncertainties and (iii) the effect of smearing.

The paper is organized as follows. Sec.~\ref{formula} depicts the theoretical background of supernova neutrinos. Going further, in Sec.~\ref{experiment}, we describe  simulation details and the experimental configuration of DUNE and T2HK, which we use in our analysis. In Sec.~\ref{result}, we present our results. We divide this section into four different subsections to discuss the points (i), (ii) and (iii) as mentioned in the previous paragraph. Finally, in Sec.~\ref{summary}, we give a summary with important conclusions.    

\section{Theoretical background}
\label{formula}

As discussed earlier, at the initial phase of the supernova burst, there is a dominance of electron-type neutrinos ($\nu_e$). They are produced by electron capture on protons and nuclei when the neutrinosphere experiences shock wave \cite{Mirizzi:2015eza}. Produced electron-type neutrinos interact with matter strongly and thus, the average energy of $\nu_e$ become less compared to other type of neutrinos ($\nu_{\mu}, \nu_{\tau}$). Further, as $\bar{\nu}_e$ has the capability to interact with matter via charged current interaction, its average energy becomes small, but not smaller than $\nu_e$. As the energy of supernova neutrinos is in the range of few MeV, non-electron type neutrinos can not interact via charged current due to high binding energy of $\mu$ and $\tau$ type leptons. They have only neutral current interaction with matter, so average energy of $\nu_x$ where $x$ can be any one of $\nu_{\mu}$ and $ \nu_{\tau}$ is highest. In the original neutrino spectra, the average energy relation for different types of neutrinos should be \cite{Dighe:1999bi}
\begin{equation}
    \langle E^0_{\nu_e} \rangle
    < \langle E^0_{\bar{\nu}_e} \rangle < \langle E^0_{\nu_x} \rangle.
\end{equation}
Flavor dependent primary neutrino spectra can be parametrized as \cite{Keil:2002in,Dasgupta:2008my,Scholberg:2012id}
\begin{equation}
    \Phi_{\nu} (E) = \mathcal{N} \left( \frac{E_{\nu}}{\langle E_{\nu} \rangle} \right)^{\alpha} e^{-(\alpha+1) \frac{E}{\langle E  _{\nu} \rangle}}\;,
\end{equation}
where $\alpha$ represents the pinching parameter, $\mathcal{N}$ is the normalisation constant expressed as
\begin{equation}
    \mathcal{N}= \frac{(\alpha+1)^{\alpha+1}}{\langle E_{\nu} \rangle \Gamma(\alpha+1)}\;,
    \label{nor}
\end{equation}
with $\Gamma$ being the Euler Gamma function. The neutrino flux ($F_{\nu}^0)$ at neutrinosphere for each flavour has the relation with $\Phi_{\nu} (E)$ 
\begin{equation}
    F_{\nu}^0 = \frac{L_{\nu}}{\langle E \rangle_{\nu}} \Phi_{\nu} (E)\;,
    \label{flux}
\end{equation}
where $L_{\nu}$ is the luminosity of the $\nu$ type neutrinos. After applying the effect of neutrino oscillation inside the star's core and surface, the modified fluxes can be grouped as \cite{Dighe:1999bi}
\begin{eqnarray}
    &&F_{\nu_e} = p F_{\nu_e}^0 + (1-p) F_{\nu_x}^0 \;,\nn\\
   && F_{\bar{\nu}_e} = \bar{p} F_{\bar{\nu}_e}^0 + (1- \bar{p} ) F_{\nu_x}^0 \;,\nn\\
&&   2F_{\nu_x} = (1-p)F_{\nu_e}^0 + (1+p) F_{\nu_x}^0\;, \nn \\
&& 2F_{\bar{\nu}_x} = (1-\bar{p})F_{\bar{\nu}_e}^0 + (1+\bar{p}) F_{\bar{\nu}_x}^0\;,
\end{eqnarray}
where $p$ and $\bar{p}$ being the survival probabilities for $\nu_e$ and $\bar{\nu}_e$ respectively. As $p$ and $\bar{p}$ are different for normal and inverted ordering of the neutrinos, in principle it is possible to determine the true mass ordering of the neutrinos by analyzing the data from supernova neutrinos. The expressions of survival probabilities for both the mass orderings are listed in table \ref{survival}. From the table, we see that the neutrino oscillation probabilities relevant for supernova neutrinos depend on the parameters $\theta_{13}$ and $\theta_{12}$.  In our study, we assume that there is no significant oscillation effect on flux of neutrinos when they travel from supernova to earth. Also, we are not taking into consideration any effect of earth-matter on the neutrino flux (more details can be found in \cite{Seadrow:2018ftp}).

\begin{table}[htbp]
    \centering
    \begin{tabular}{|c|c|c|}
    \hline
    \rowcolor{citrine!10}
        Ordering & $p$ & $\bar{p}$  \\
        \hline
        \rowcolor{babypink!50}
        Normal & $\sin^2 \theta_{13}$ & $\cos^2 \theta_{12} \cos^2 \theta_{13}$  \\
        \hline
        \rowcolor{aureolin!30}
         Inverted & $\sin^2 \theta_{12} \cos^2 \theta_{13}$  &  $\sin^2 \theta_{13}$\\
        \hline
    \end{tabular}
    \caption{Survival probability expressions of neutrino ($p$) and antineutrino ($\bar{p}$) fluxes for two cases: normal ordering and inverted ordering.}
    \label{survival}
\end{table}

 Here it is important to note that apart from the oscillations induced by the above mentioned Mikheyev-Smirnov-Wolfenstein (MSW) effect, there can be other effects which may affect the flavour oscillations of the supernova neutrinos. One of these effects is collective transitions arising from the neutrino self interactions. These can be either slow \cite{Mirizzi:2015eza,Chakraborty:2016yeg,Horiuchi:2018ofe} or fast \cite{Tamborra:2020cul}, according to the size of the timescale for their development. The slow collective effect can lead to either a spectral swaps which happens only for inverted ordering at 8 MeV \cite{Chiu:2013dya} or there can be multiple spectral splits at different energies and also for normal ordering \cite{Dasgupta:2009mg,Friedland:2010sc}. It is now believed that spectral swaps due to slow effects are suppressed in realistic situations \cite{Sarikas:2011am,Chakraborty:2011nf}. Further, several recent studies have found that the fast instabilities lead to flavour equilibrium prior to the MSW region \cite{Bhattacharyya:2022eed,Wu:2021uvt,Richers:2021nbx,Richers:2021xtf,Sigl:2021tmj}. From the above discussion we understand that study of collective effects is an active area of research and their effect on neutrino flavour conversions are yet to be understood fully. Nevertheless, a full multi-angle study of neutrino self-interactions showed that the energy-dependent modifications of the spectrum would get smeared out when considering the post-bounce time-integrated spectrum and corrections are expected to be small \cite{Lunardini:2012ne}. Therefore, in our analysis we will ignore the transitions induced by collective effects.

\begin{figure}
    \centering
    \includegraphics[scale=0.9]{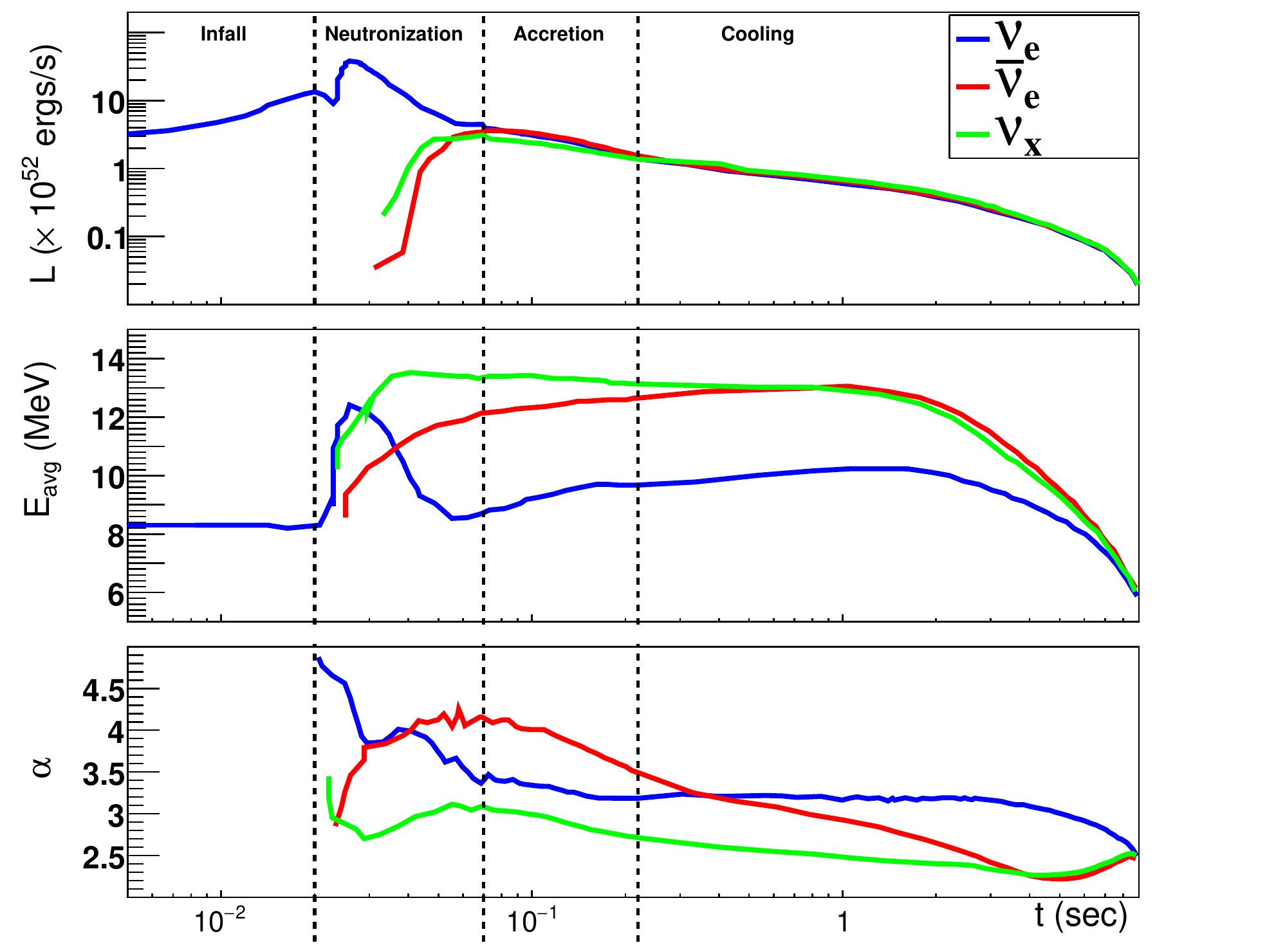}
    \caption{Time dependent flux parameters as a function of time (in seconds) for Garching model\cite{Hudepohl:2009tyy}. In the top panel, luminosity of neutrino flux is plotted against the supernova evolution time for three different neutrino flavours: $\nu_e$, $\bar{\nu}_e$ and $\nu_x$, where $x$ represent any one of the non-electron neutrino flavour. Middle panel shows the variation of average energy of each neutrino flavour with time. Bottom panel is for the variation of pinching parameter ($\alpha$) for different flavours over time. Pinching parameter of $\nu_e$ in infall region is fixed at 4.87353. In each panel, blue, red and green curves are for $\nu_e, \bar{\nu}_e$ and $\nu_x$ respectively. This figure is made by the software SNOwGLoBES \cite{github}}.
    \label{time-gar}
\end{figure}

\begin{figure}[htbp]
%\hspace{-1.0cm}
    \subfloat[]{\includegraphics[scale=0.5]{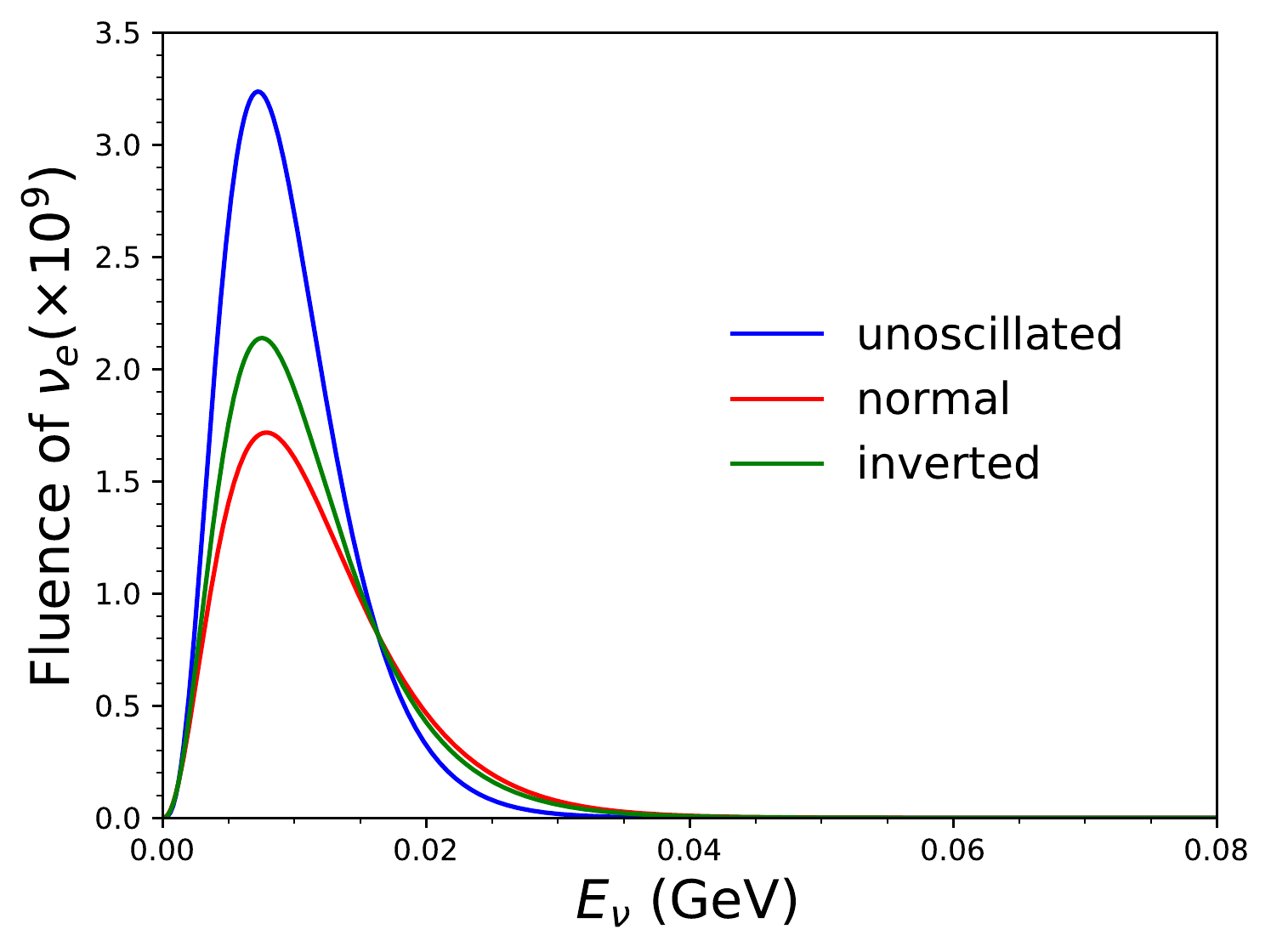}\label{flux-nu-a}}
\subfloat[]{\includegraphics[scale=0.5]{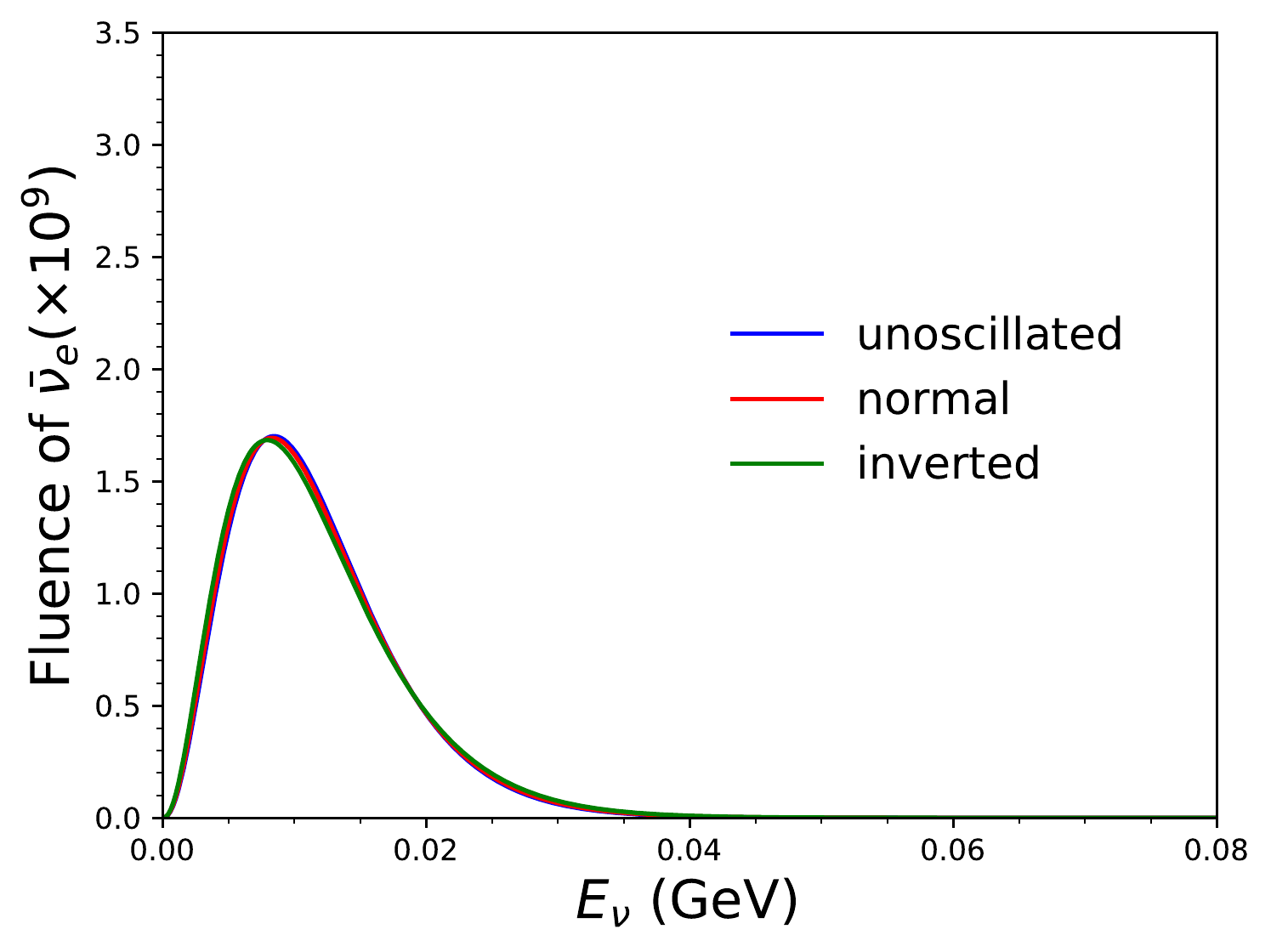}\label{flux-anu-a}}\\
 \subfloat[]{\includegraphics[scale=0.5]{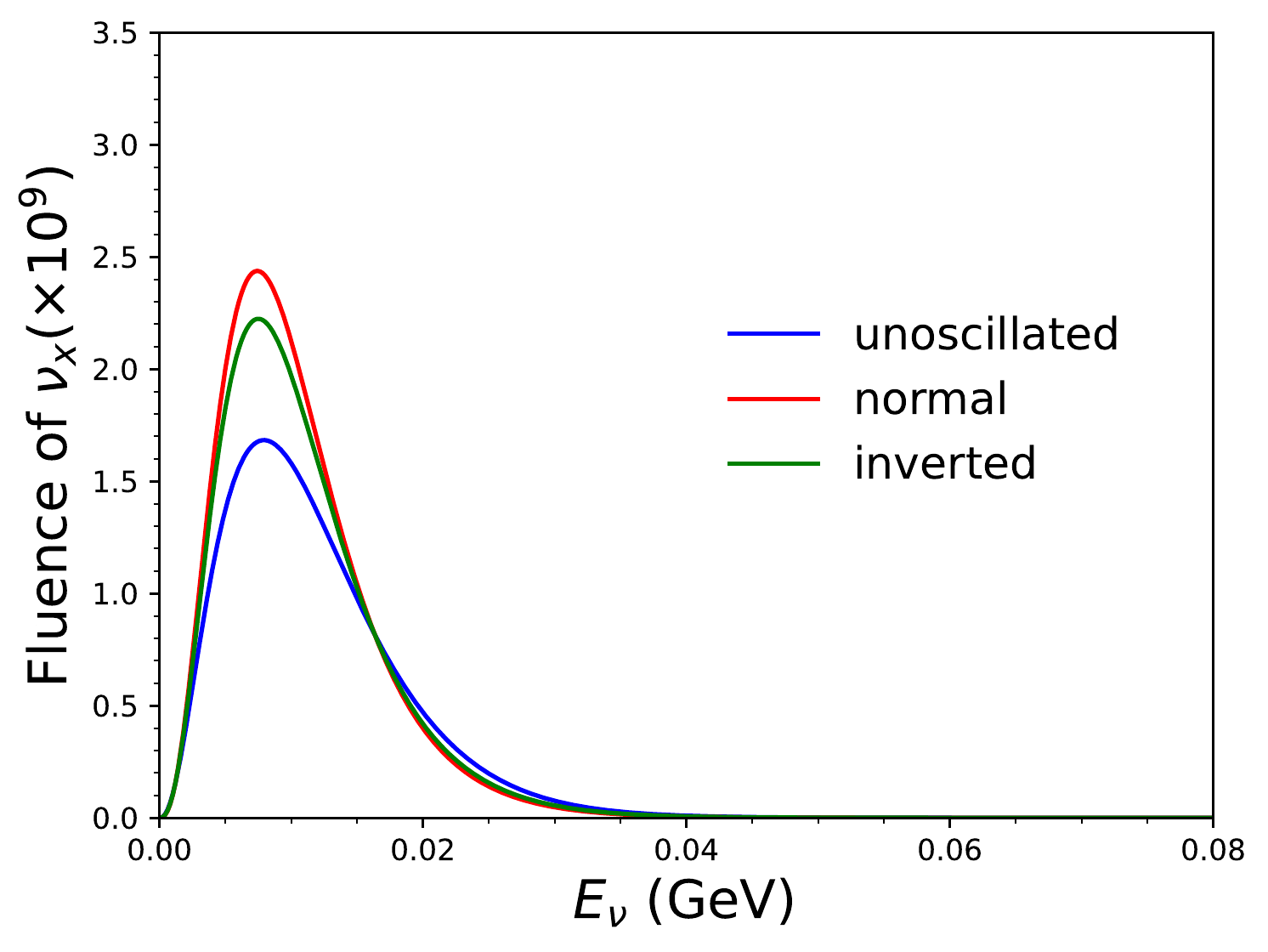} \label{flux-nux-a}}
 \subfloat[]{\includegraphics[scale=0.5]{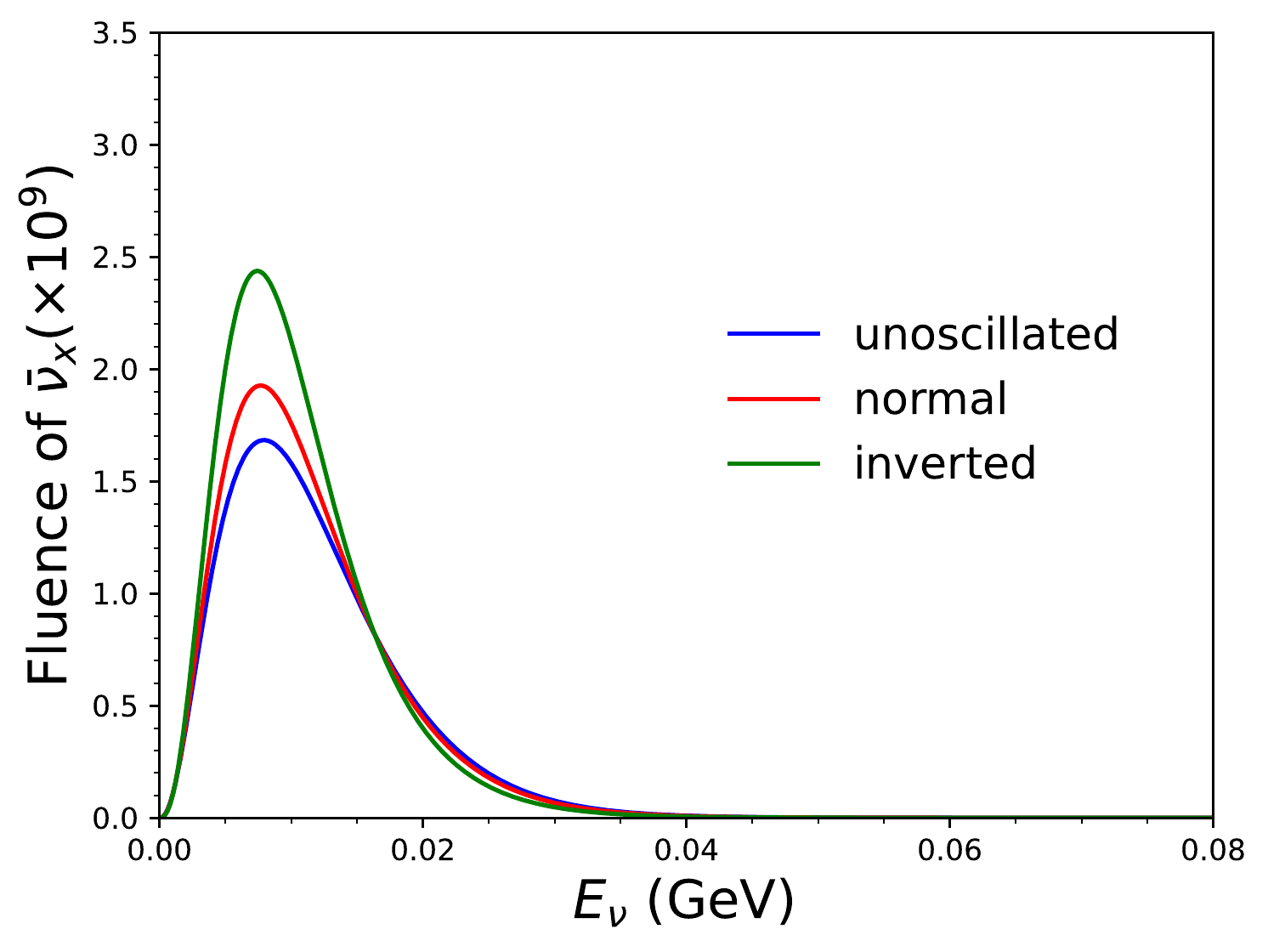} \label{flux-nuxbar-a}}
    \caption{(\textbf{\ref{flux-nu-a})} Fluence of $\nu_e$ vs energy (in GeV), \textbf{(\ref{flux-anu-a})} Fluence of $\bar{\nu}_e$ vs energy (in GeV) and \textbf{(\ref{flux-nux-a})} Fluence of $\nu_x$ vs energy (in GeV) for Garching model. In each panel, blue, red and green curves are for unoscillated, normal ordering and inverted ordering respectively. }
    \label{fluxes-nu}
\end{figure}

In Fig.~\ref{time-gar}, we have shown the variation of time dependent flux parameters with respect to supernova evolution time for Garching electron-capture supernova model \cite{Hudepohl:2009tyy}. As mentioned earlier, this flux model has been considered in the analysis by DUNE collaboration \cite{DUNE:2020zfm} and we have explicitly checked that our event numbers indeed match with their result. In the top panel, we have plotted the luminosity of the flux as a function of time. In the middle panel, average energy of three neutrino flavours has been shown. Finally, in the bottom panel, the pinching parameter ($\alpha$) is depicted as a function of time. In each panel, blue, red and green coloured curves are for $\nu_e$, $\bar{\nu}_e$ and $\nu_x$ respectively. Also, in each panel, three stages of explosion have been mentioned: neutronization, accretion and cooling. The time of 0.02 seconds to 50 ms is the neutronization burst era. Similarly, from 50 ms to 200 ms is the accretion period. Finally, from 0.2 sec to 9 seconds is known as the proto-neutron-star cooling period.  

We have calculated the time integrated flux by folding the information of Fig.~\ref{time-gar} into Eq.~\ref{flux}. We have presented the time integrated fluxes i.e., the fluences in Fig.~\ref{fluxes-nu}. Panel \ref{flux-nu-a} shows the fluence of electron-type neutrino with respect to neutrino energy (in GeV), whereas panel \ref{flux-anu-a} is the fluence of $\bar{\nu}_e$ as a function of neutrino energy. Panel \ref{flux-nux-a}  (\ref{flux-nuxbar-a}) depicts the same for $\nu_x$ ($\bar{\nu}_x$) as a function of neutrino energy. 
.
In each panel, blue, red and green curves show the fluence of $\nu_e, \bar{\nu}_e$ and $\nu_x$ for three different cases, i.e. unoscillated, normal ordering and inverted ordering respectively. We take the values of $\theta_{12}$ and $\theta_{13}$ as $33.41^\circ$ and $8.58^\circ$ respectively \cite{Esteban:2020cvm}. 
For  the $\nu_e$, $\nu_x$  and $\bar{\nu}_x$ cases, we see that the oscillated event spectrum is different from the unoscillated in spectrum. The difference is observed mainly at the peak.
In the panel \ref{flux-nu-a}, the unoscillated flux is greater than the oscillated flux whereas in \ref{flux-nux-a} and \ref{flux-nuxbar-a} panels, it is opposite. The difference between unoscillated, normal and inverted ordering is not clearly distinguishable in \ref{flux-anu-a} panel. 
In the panels \ref{flux-nu-a} and \ref{flux-nuxbar-a}, the flux for the inverted ordering is higher as compared to the flux for the normal ordering. This is opposite for the panel \ref{flux-nux-a}. As the fluxes for the normal ordering and inverted ordering are different for panels \ref{flux-nu-a} and \ref{flux-anu-a}, it is possible to discriminate between the two orderings by analyzing the oscillated $\nu_e$ and $\bar{\nu}_e$ event rates from the supernova neutrinos. 
From panel \ref{flux-nux-a} and panel \ref{flux-nuxbar-a}, we notice that there is also a separation between the fluxes for normal ordering and inverted ordering for $\nu_x$ and $\bar{\nu}_x$. But these events can be observed in the detector only via neutral current interactions. As the neutral current interaction will also include the contribution from $\nu_e$ and $\bar{\nu}_e$ events, the total number of neutral current events observed at the detector will be same as the number of neutrinos produced at the supernova. Therefore, for three flavour, it will not be possible to study the effect of flavour conversion by studying neutral current events at the detector and hence one cannot determine the neutrino mass ordering by studying the same.

\section{Experiment and simulation details}
\label{experiment}

For the analysis of the supernova neutrinos, we have used SNOwGLoBES (Supernova Neutrino Observatories with GLoBES)\cite{github} software. This is a software based on GLoBES (General Long Baseline Experiment Simulator) \cite{Huber:2004ka, Huber:2007ji} package specially designed for supernova neutrinos. SNOwGLoBES is an event rate calculator by the use of input fluxes, cross-section and detector responses. 

In our analysis, we have considered two future neutrino detectors: the  Liquid Argon Time Projection Chamber (LArTPC) far detector (FD) of DUNE and the  water Cherenkov FD of T2HK. Far detector of DUNE (Deep Underground Neutrino Experiment) is located 1.5 km underground at the Sanford Underground Research Facility in South Dakota. DUNE has four liquid Argon time projection chamber modules, each with 10 kt fiducial volume. Main advantage of using liquid Argon as detector material is that, it has very good sensitivity to interact with low energy neutrinos \cite{ArgoNeuT:2016wjb}. We take Gaussian energy resolution of $20\%$ as mentioned in Ref.~\cite{DUNE:2020zfm}. Main interaction of low energy supernova neutrinos with DUNE detector is $\nu_e-^{40}{\rm Ar}$ charged current interaction:
\begin{equation}
    \nu_e + ^{40}{\rm Ar} \rightarrow e^- + ^{40}K^* \;.
\end{equation}
For our convenience, we call the main interaction channel mode as ``channel A". Having excellent detection capability of low energy $\nu_e$ components, DUNE can be considered as first ever supernova neutrino detector to probe $\nu_e$ signals, while most of the currently running or future detectors are sensitive to antineutrino signals. DUNE will also be able to detect $\bar{\nu}_e$ as its second dominant detection channel through the interaction: 
\begin{equation}
    \bar{\nu}_e + ^{40}{\rm Ar} \rightarrow e^+ + ^{40}{\rm Cl}^*\;.
    \label{channelb}
\end{equation}
We call this interaction channel as ``channel B".  Elastic scattering of neutrinos with electron can provide accurate position of supernova by seeing the direction of momentum of scattered electron. DUNE can detect neutrino signal through elastic scattering with electron by the interaction channel
\begin{equation}
    \nu_e + e^- \rightarrow e^- + \nu_e\;,
\end{equation}
and we call this interaction channel as ``channel C".

Hyper-Kamiokande (HK) is a future water Cherenkov detector which is going to be installed in Japan, approximately 8 km south to Super-Kamiokande. Future long-baseline experiment with HK as far detector 295 km from the source at J-PARC is known as T2HK (Tokai to Hyper-Kamiokande). HK will be composed of two very large cylindrical detectors each with height of 71 m and diameter of 68 m, filled with 187 kt fiducial volume of ultra-pure water. Coverage of photo-multiplier tube (PMT) has been not yet decided for the actual installation of the detector, but for our study, we take $40 \%$ photocoverage with the new 50 cm PMT model~\cite{Hyper-Kamiokande:2021frf}. For this experiment, we take Gaussian energy resolution of $18\%$. Having ultra-pure water as the detector material, the main channel for supernova neutrino detection is inverse beta decay (IBD). In IBD, incoming $\bar{\nu}_e$ interact with proton of the water molecule to form neutron and positron:
\begin{equation}
    \bar{\nu}_e + p \rightarrow n + e^+.
 \end{equation}
We call the main interaction channel, IBD as ``channel A" for T2HK experiment. In HK detector, almost $90 \%$ events are IBD. In the presence of $^{16} O$ molecule, second leading interaction channel is
\begin{equation}
    \bar{\nu}_e + ^{16}O \rightarrow e^+ +  ^{16}N^*.
    \label{channel-b}
\end{equation}
This channel is referred as ``channel B" in our study for T2HK experiment.
Another promising detection channel is the interaction of electron-neutrino with electron. We call it as ``channel C",
\begin{equation}
    \nu_e + e^- \rightarrow e^- + \nu_e.
\end{equation}
All the above mentioned channels are listed in table \ref{channel}.

For our analysis, we take the energy range of supernova neutrinos from 0.5 MeV to 100 MeV with 200 true energy bins and 200 sampling bins for both  the experiments. In our analysis, we do not consider any background. This is because, for Galactic supernova burst, the rate of backgrounds in current and future experiments are very low. Background for supernova neutrinos can come from radioactivity, cosmic ray, reactor $\bar{\nu}_e$, solar $\nu_e$ etc. Even some of the backgrounds can come from low energy atmospheric neutrinos and antineutrinos. Fortunately, most of these can be suppressed by taking the detector underground. More details on background for supernova neutrinos can be found in Ref.~\cite{Scholberg:2012id}.

\begin{table}[htbp]
    \centering
    \begin{tabular}{|c|c|c|c|}
    \hline
    \rowcolor{citrine!10}
        Experiment  &  Channel-A & Channel-B & Channel-C \\
        \hline
        \rowcolor{chartreuse(web)!30}
        DUNE &  $\nu_e-^{40}\rm Ar $  &  $\bar{\nu}_e-^{40}\rm Ar $ & $\nu_e-e$ \\
        \hline
        \rowcolor{candypink!20}
        T2HK & IBD & $\bar{\nu}_e-^{16}O$ & $\nu_e-e$\\
        \hline
    \end{tabular}
    \caption{Different interaction modes for two experiments, DUNE and T2HK}
    \label{channel}
\end{table}

For the estimation of neutrino mass ordering sensitivity, first we calculate the expected number of events corresponding to the channels that we mention above. Then we define a Poisson log-likelihood formula to calculate the statistical $\chi^2$,
\begin{equation}
    \chi^2_{\rm stat} = 2 \sum_{i=1}^n \left[N_i^{\rm test} - N_i^{\rm true} - N_i^{\rm true} \rm{log} \left( \frac{N_i^{ \rm test}}{N_i^{\rm true}} \right) \right].
    \label{chi}
\end{equation}
In the above formula, we take event rates of normal mass ordering as true event rates and $N_i^{\rm test}$ is the event rates of inverted ordering with $i$ as number of energy bins.

To incorporate the effect of systematic errors in our calculation, we have taken two types of systematic errors: normalisation error and energy calibration error.
If we assume a systematic error of 5\% for both normlisation and energy calibration, then the modified expression of $N_{i}^{\rm test}$ becomes
 \begin{equation}
    N_i^{\rm test} \rightarrow N_i^{\rm test} [( 1+ 0.05 \zeta_1 ) + 0.05 \zeta_2(E_i^{\prime} - \bar{E^{\prime}})/(E^{\prime}_{\rm max}-E^{\prime}_{\rm min})]
    \label{error}
\end{equation}
 where $\zeta_1$ and $\zeta_2$ are the pull variables responsible for normalisation and energy calibration errors respectively. $E^{\prime}_{\rm max}$ and $E^{\prime}_{\rm min}$ are the maximum and minimum energy range of the event spectrum respectively. $\bar{E}^{\prime}= \frac{1}{2}(E^{\prime}_{\rm max}+E^{\prime}_{\rm min})$ is the median of the energy interval and $E_i^{\prime}$ is the reconstructed energy of $i$th bin.
Thus total sensitivity (stat + sys) can be grouped as
\begin{equation}
    \chi^2_{ \rm stat+sys} = \chi^2_{\rm stat} + \zeta_1^2 +\zeta_2^2 \;.
    \label{chi-sys}
\end{equation}
In our work, we use eq. \ref{chi} for the calculation of sensitivity without systematics and eq. \ref{chi-sys} for the calculation of $\chi^2$ with systematics.  We will discuss the effects of these two types of systematic errors on mass ordering sensitivity in Sec.~\ref{effect:sys}. In the following section, we will present the analysis of mass hierarchy sensitivity in the context of DUNE and T2HK experiments with the above mentioned three different interaction channels.

\section{Results}
\label{result}

In this section, we present our results. We divide this section into four subsections. In Subsec.~\ref{event-sec} we present the event rates for different channels for T2HK and DUNE. Subsec.~\ref{effect:event} demonstrates the neutrino mass ordering sensitivity corresponding to different channels, while ~\ref{effect:sys} shows the effect of systematics in the measurement of neutrino mass ordering sensitivity. Finally, Subsec. \ref{effect:smear} depicts the effect of smearing for the estimation of mass ordering $\chi^2$.

\subsection{Event-rates}
\label{event-sec}

\begin{figure}[htbp]
%    \centering
    \hspace{-1.2cm}
   \subfloat[]{\includegraphics[scale=0.37]{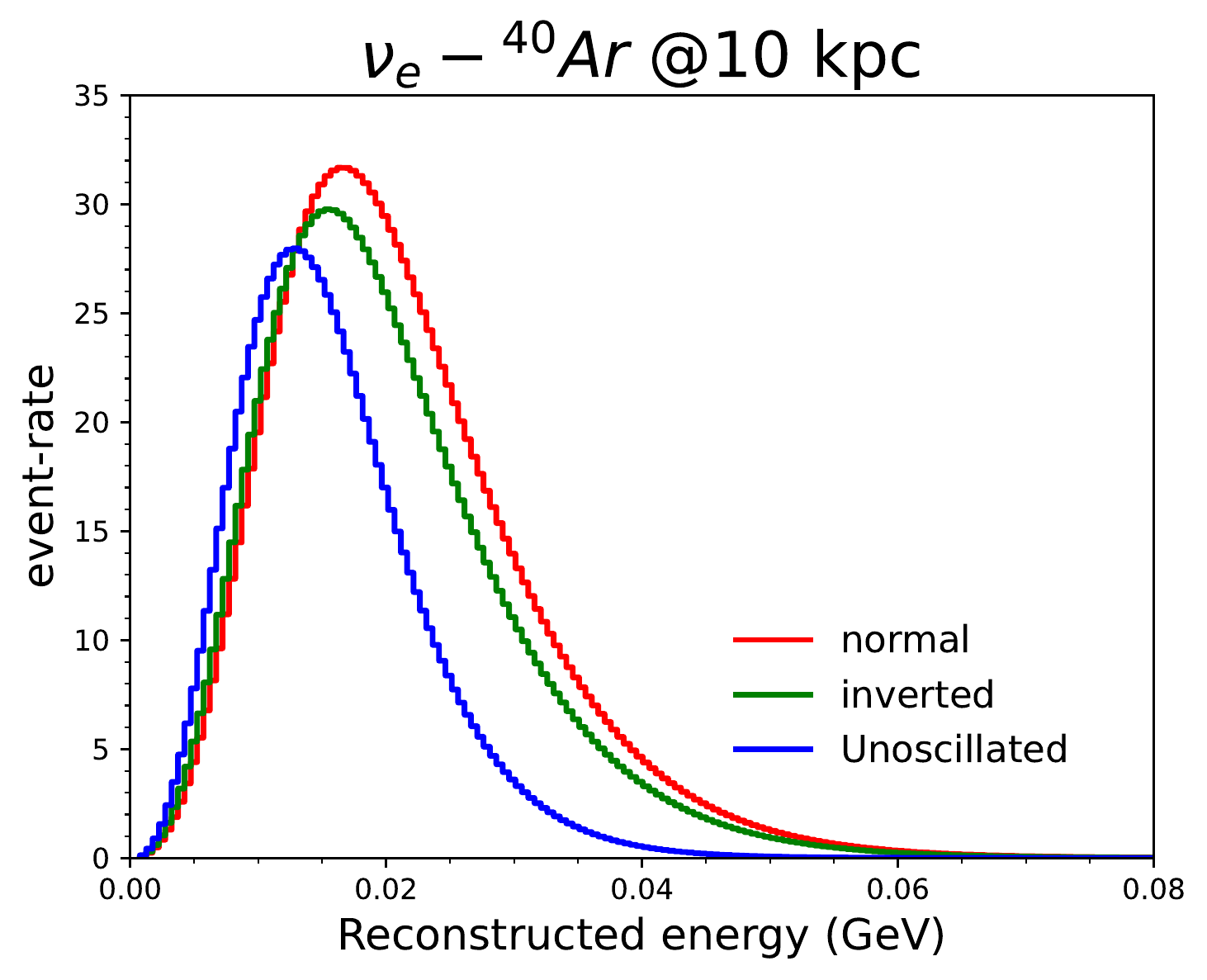}\label{dune-cha}}
    \subfloat[]{\includegraphics[scale=0.37]{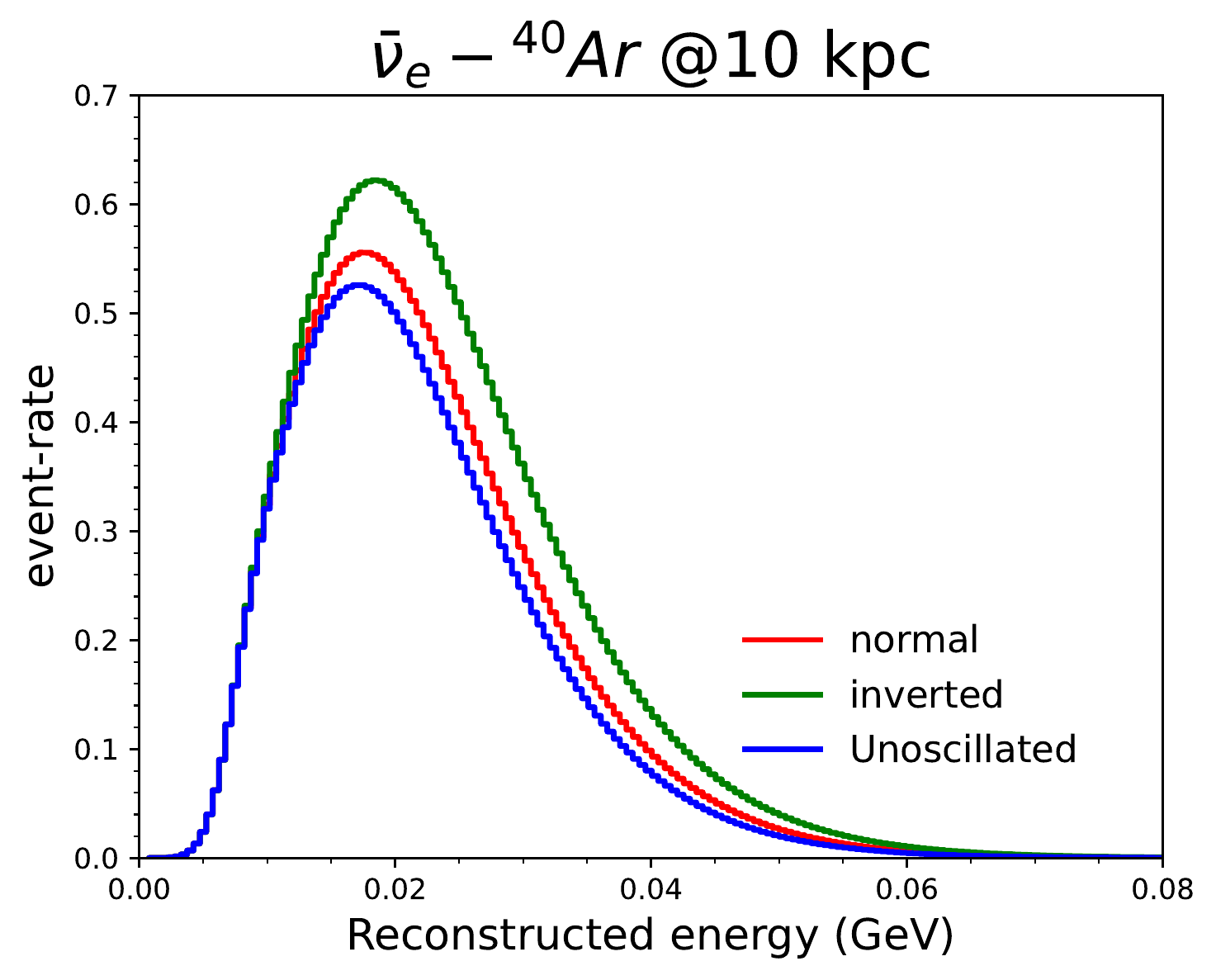}\label{dune-chb}}
    \subfloat[]{\includegraphics[scale=0.37]{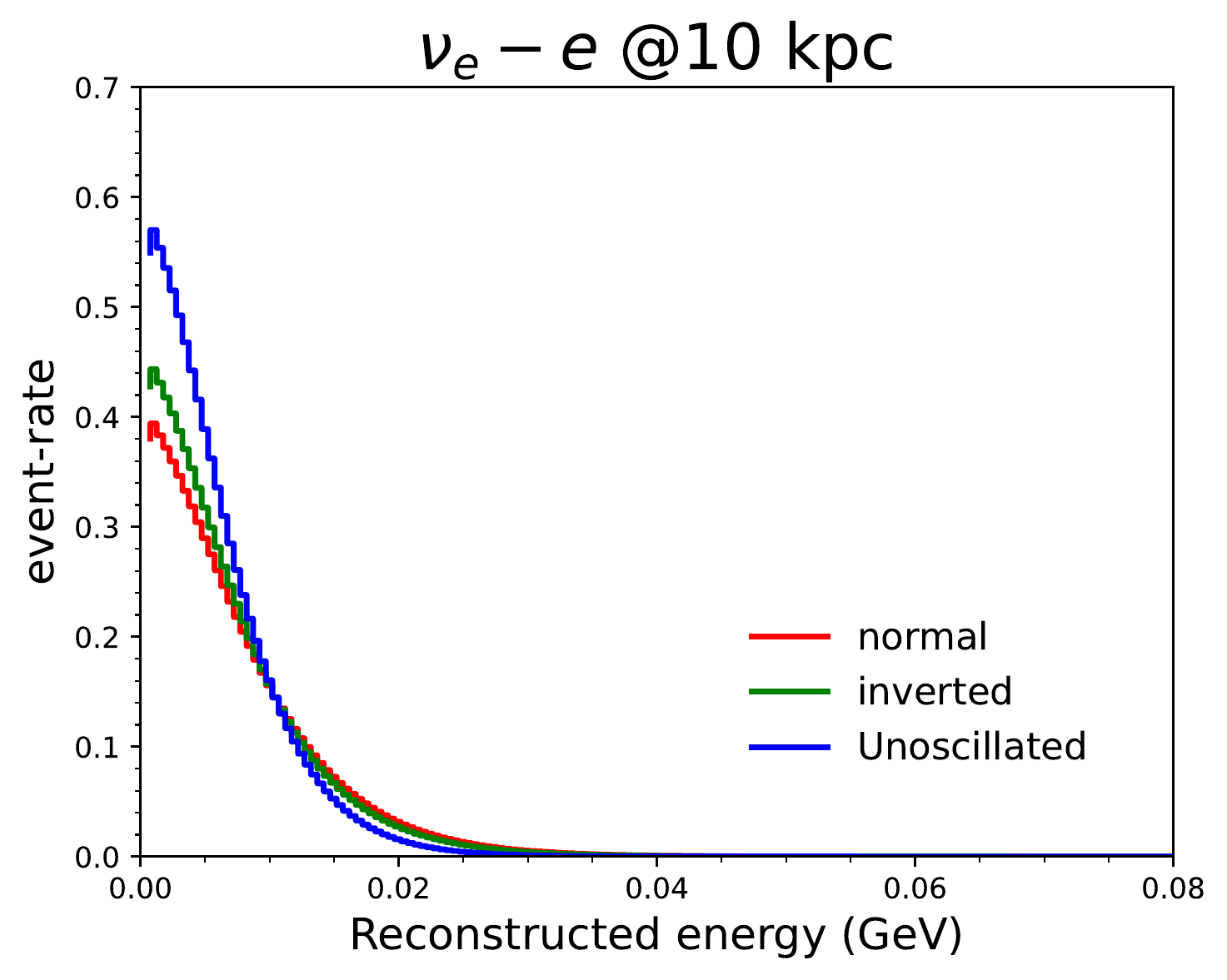}\label{dune-chc}} \\
    \hspace{-1.2 true cm}
    \subfloat[]{\includegraphics[scale=0.37]{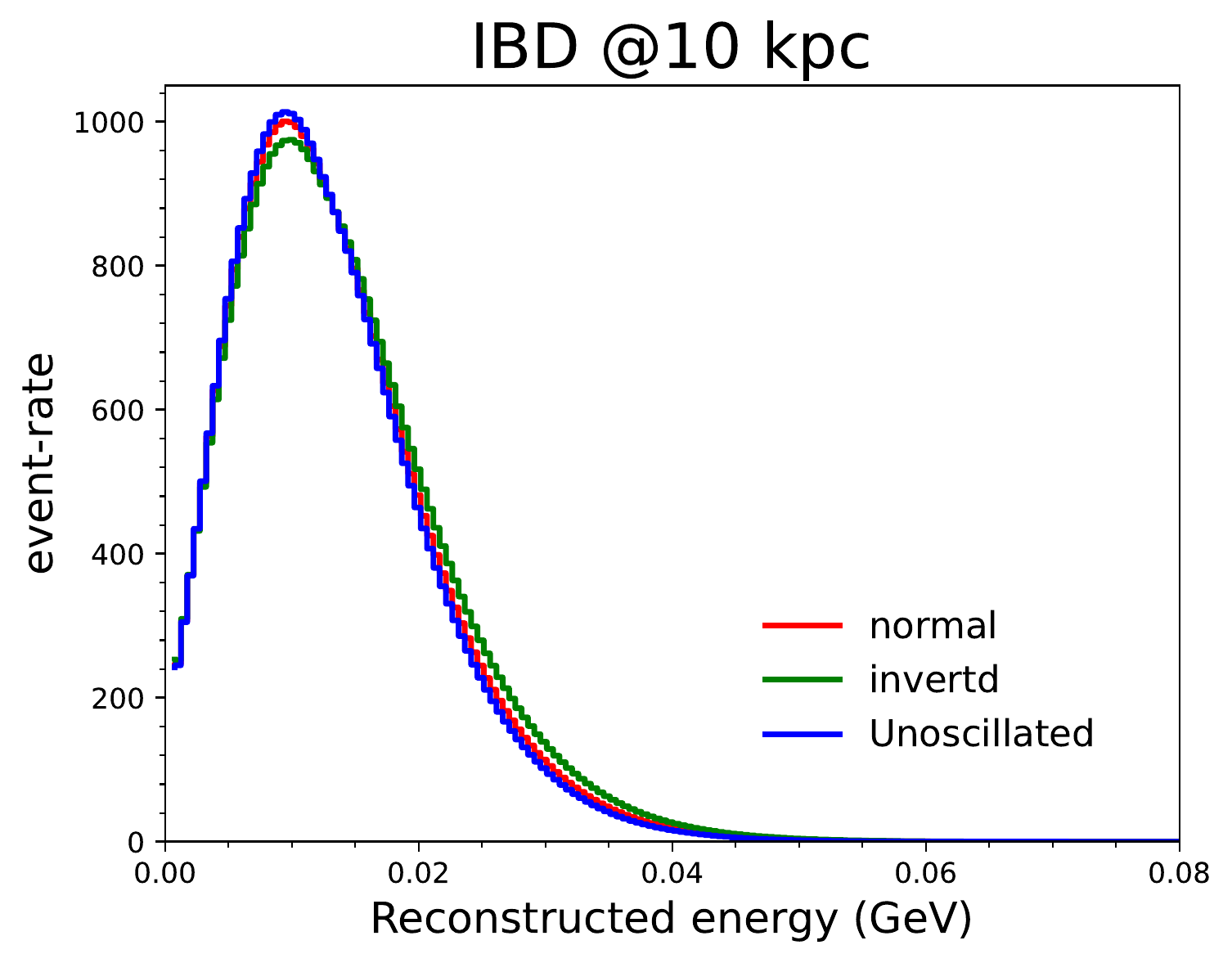}\label{t2hk-cha}}
    \subfloat[]{\includegraphics[scale=0.37]{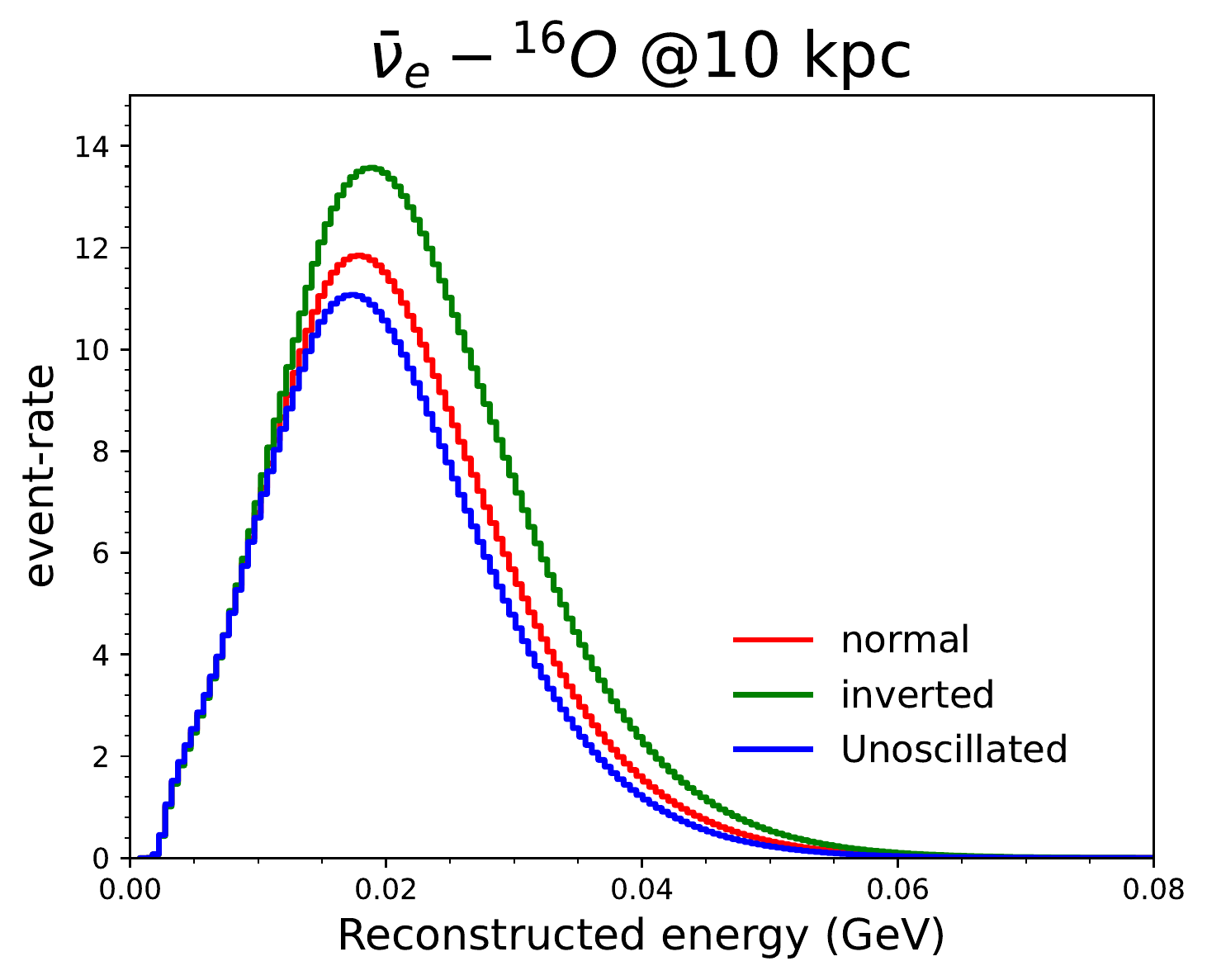}\label{t2hk-chb}}
    \subfloat[]{\includegraphics[scale=0.37]{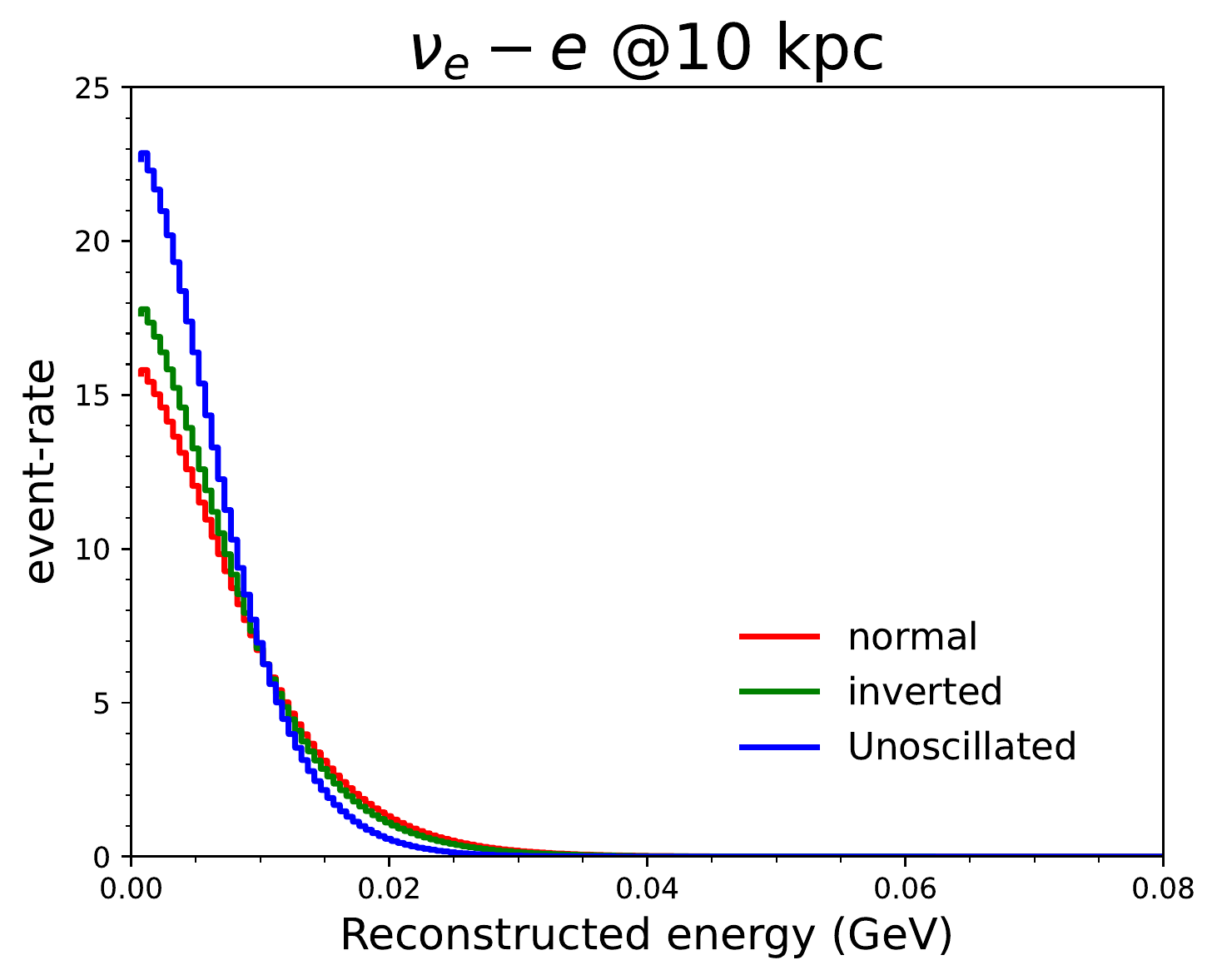}\label{t2hk-chc}}
    \caption{Event-rates vs reconstructed energy for DUNE and T2HK in three different cases: unoscillated (blue), normal ordering (red) and inverted ordering (green). The distance of the supernova is assumed to be 10 kpc. The panels \textbf{(\ref{dune-cha})} Channel A, \textbf{(\ref{dune-chb})} Channel B and \textbf{(\ref{dune-chc})} Channel C are for DUNE and \textbf{(\ref{t2hk-cha})} Channel A, \textbf{(\ref{t2hk-chb})} Channel B, \textbf{(\ref{t2hk-chc})} Channel C are for T2HK.}
    \label{event}
\end{figure}

In table~\ref{event-table}, we summarize the total supernova neutrino events to be detected at DUNE and T2HK, corresponding to the different channels which we discussed in Sec~\ref{experiment}. For this purpose, we assume the distance of the supernova to be 10 kpc. In this table, the orange shaded rows show event numbers for three different channels and their combinations in DUNE experiment with three cases: unoscillated, normal ordering and inverted ordering. From this table, we see that the event numbers in channel B is roughly one order of magnitude less than channel A and the events in channel C is one order of magnitude less than channel B. As the event numbers in channel B and C are very less, we conclude that channel A will contribute largely to the sensitivity of neutrino mass ordering as compared to channel B and channel C. To understand the variation of events with respect to energy, in the top row of Fig. \ref{event}, we show the event spectrum for DUNE as a function of reconstructed energy. Panel~\ref{dune-cha} shows the event rates for channel A, panel~\ref{dune-chb} is for channel B and panel~\ref{dune-chc} for channel C. In each panel, we demonstrate three cases: blue curve is for unoscillated condition whereas red and green curves are for normal ordering and inverted ordering, respectively. Unlike panel \ref{flux-nu-a} and \ref{flux-anu-a}, the event spectra of panel \ref{dune-cha} and \ref{dune-chb} have  unoscillated spectra lower than the oscillated spectra. However, in channel A, event rate for normal ordering is higher than the event rate for inverted ordering and it is opposite in channel B. This feature in these two panels are different as compared to the flux spectra in Fig.~\ref{fluxes-nu} where we have the fluence for normal ordering lower than the fluence for inverted ordering for $\nu_e$ and for $\bar{\nu}_e$, the difference between normal and inverted ordering is marginal. We have checked that this happens because of the energy dependence of the corresponding cross-sections in channel A and channel B which are folded with the fluxes in order to calculate the event rates. Moving further, for channel C in DUNE, the shape of the event spectrum is very different from rest of the panels. We checked that the reason for this shape is due to the effect of energy smearing. In absence of energy smearing, event rate spectrum for channel C is similar to the other panels.

\begin{table}[htbp]
  \centering
  \begin{tabular}{|l|c|c|c|c|c|c|c|}
    \hline
    \rowcolor{citrine!10}
    Setup & Oscillation  & \multicolumn{5}{c|}{Channels}  \\
    & mode & \multicolumn{5}{c|}{(no. of events)}\\
    \hline 
    \rowcolor{citrine!10}
    &  & Channel A ~&~ Channel B ~& Channel C & Channel (B+C)  & Channel (A+B+C) \\
    \hline
    \rowcolor{antiquewhite!90}
       &  unoscillated  &  902.2  & 22.94  & 8.812 &   31.75 & 933.9 \\
    \cline{2-5}
    \rowcolor{antiquewhite!90}
    DUNE & NO & 1381 & 25.07 & 7.497  &   32.56  & 1414 \\
    \cline{2-5}
    \rowcolor{antiquewhite!90}
    &   IO  & 1247 & 29.46 & 7.866 &  37.32  &   1284\\
    \hline 
    \rowcolor{aqua!30}
      &  unoscillated  &  35193  & 474.48 & 366.18 &   840.66  &  36034\\
      \cline{2-5}
       \rowcolor{aqua!30}
     T2HK  &   NO   & 35649 &   523.13 &  311.53 & 834.66  &   36484\\
      \cline{2-5}
       \rowcolor{aqua!30}
      &   IO & 36591 & 623.71 & 326.86 &   950.57  &   37541\\
    \hline
     \end{tabular}
     \caption{Event numbers for different channels in two experiments : DUNE and T2HK at 10 kpc supernova distance. In the table, ``NO" and ``IO" refers to ``normal ordering" and ``inverted ordering" respectively.}
     \label{event-table}
     \end{table}

Skyblue shaded region of table \ref{event-table} shows the event number for three different channels for T2HK experiment. Here we see that the event numbers in channel B and C are approximately two order of magnitude less than channel A. However, as the event numbers of channel B and C are also large, all of the three channels can have significant contribution in the measurement of mass ordering sensitivity. To see the variation of the events with energy, in the bottom panel of Fig. \ref{event}, we show the event spectrum for T2HK as a function of reconstructed energy similar to that of DUNE. In the left and middle panels, the unoscillated spectra are lower than oscillated spectra whereas in the right panel, the unoscillated spectrum is higher than the oscillated spectrum. The event spectrum corresponding to normal ordering is higher as compared to the events for the inverted ordering for channel A, though the difference is very small. For channel B and C, the nature of event spectrum is opposite to channel A. As in DUNE, we checked that the differences between the flux spectra and the event spectra for T2HK arise because of the different energy dependence of the cross-sections. Similar to DUNE, for channel C, the shape of the event spectrum is very different from rest of the panels. Here also the  reason for this shape is the effect of energy smearing. When there is no smearing of energy, event rate spectrum for channel C is identical to the other panels.

\begin{figure}[htbp]
    \centering
    \includegraphics[height=70mm, width=90mm]{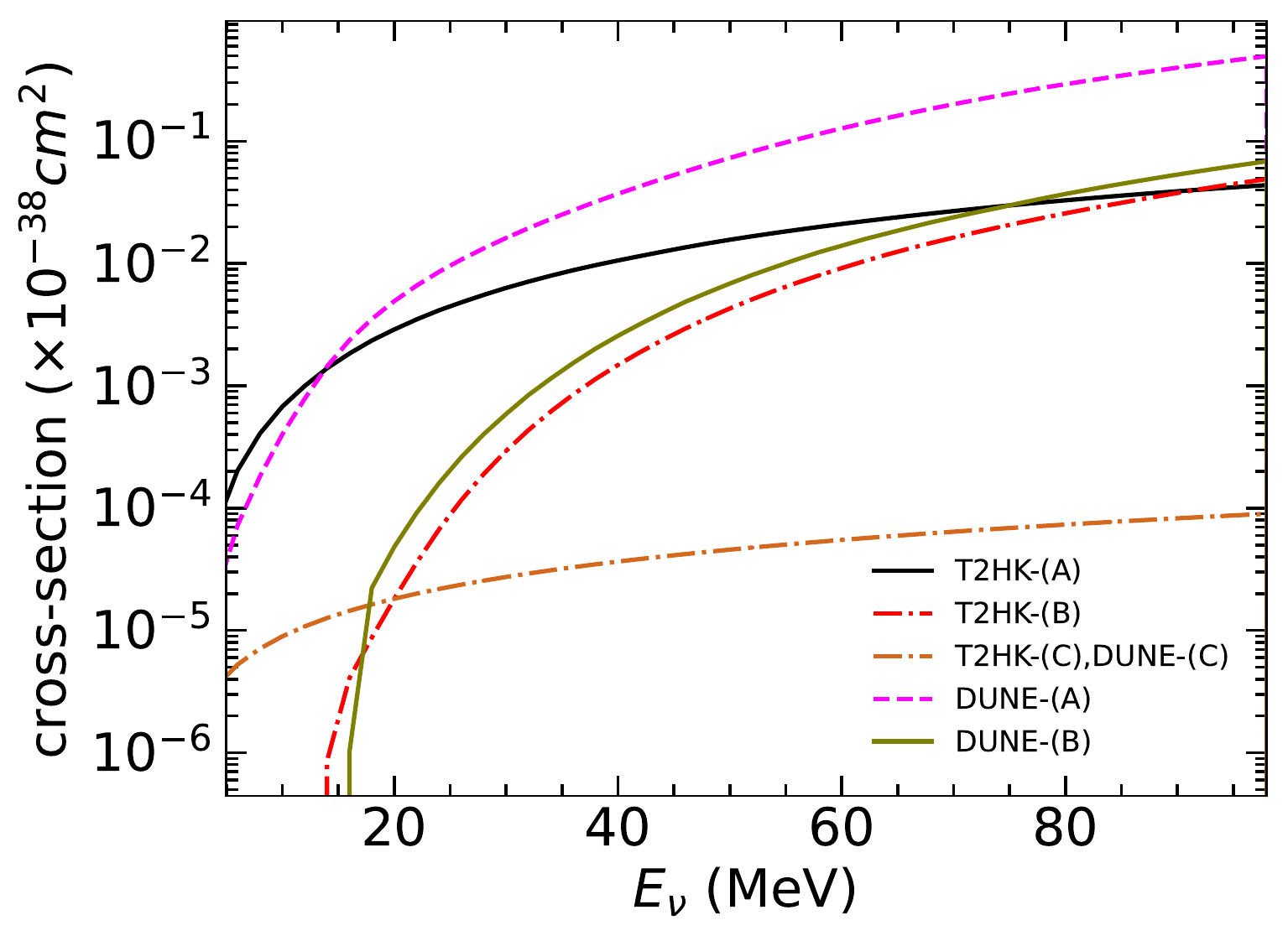}
    \caption{Cross-section of different channels for DUNE and T2HK experiments as function of neutrino energy (in MeV).}
    \label{cross-section}
\end{figure}

From the table~\ref{event-table} and the figure~\ref{event}, we see that the number of events in T2HK are higher  compared to the number of events in DUNE. We verified that, this is because of the large detector volume of T2HK, rather the effect of different cross-sections which are responsible for the detection of supernova neutrinos in different channels in both these experiments. This can be seen from Fig.~\ref{cross-section}, where we plot the cross-section for different processes as a function of energy. This figure contains five curves, showing the relevant cross-sections for DUNE and T2HK. From fig. \ref{cross-section}, we can see that the cross-section of IBD interaction (channel A of T2HK) is lower than the cross-section of $\nu_e-^{40}{\rm Ar}$ interaction (channel A of DUNE). Also, cross-section for channel B of DUNE is higher than the cross-section for channel B of T2HK. Despite these facts, we see that the number of events in T2HK for channel A and B are higher than DUNE.
Channel C of both the experiments are same ($\nu_e-e$) and thus the cross section for these two channels are same.

\subsection{Effect of different channels}
\label{effect:event}

\begin{figure}[htbp]
     \centering
     \subfloat[]{\includegraphics[height=60mm, width=80mm]{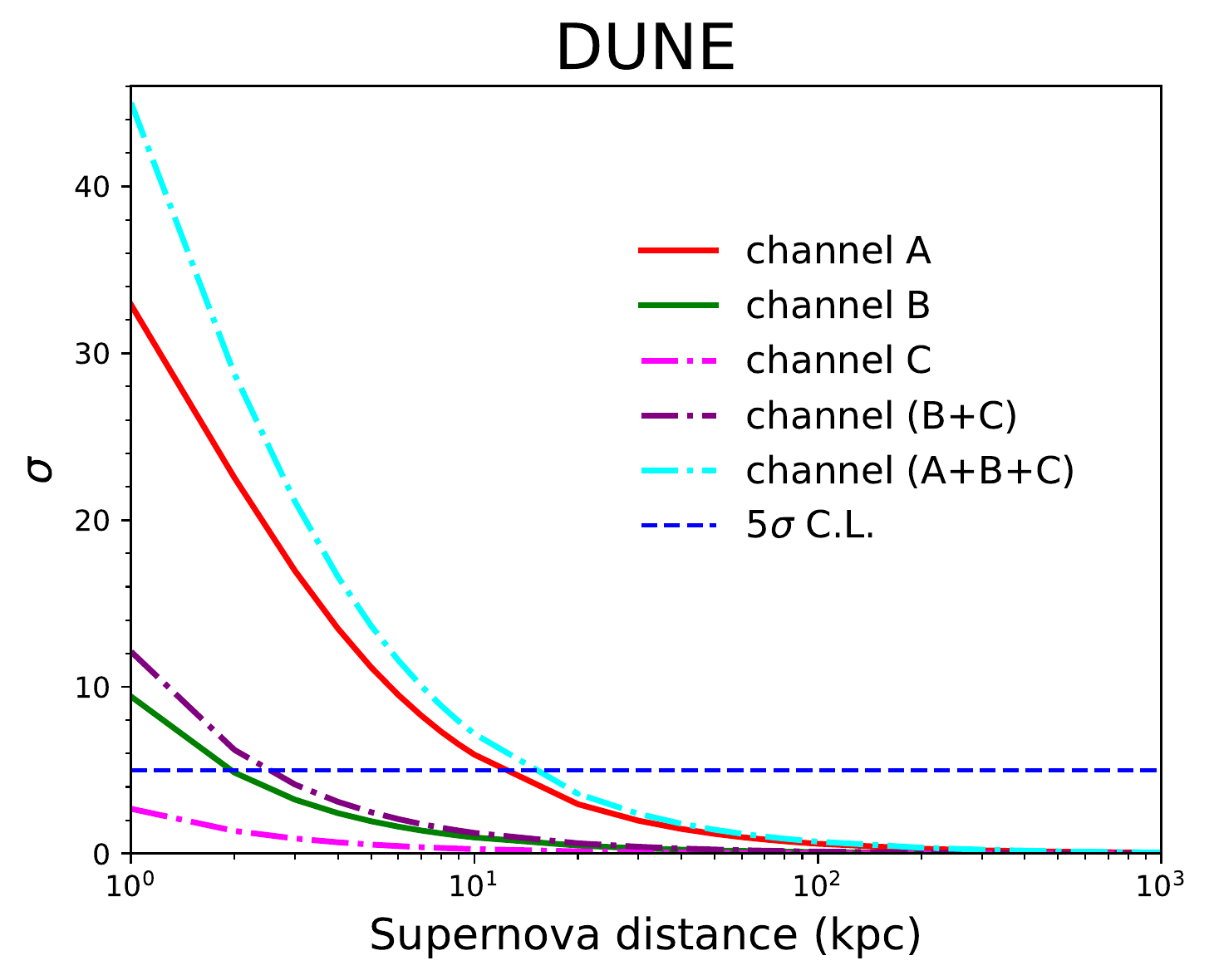}\label{channel-distance-dune}}
     \subfloat[]{\includegraphics[height=60mm, width=80mm]{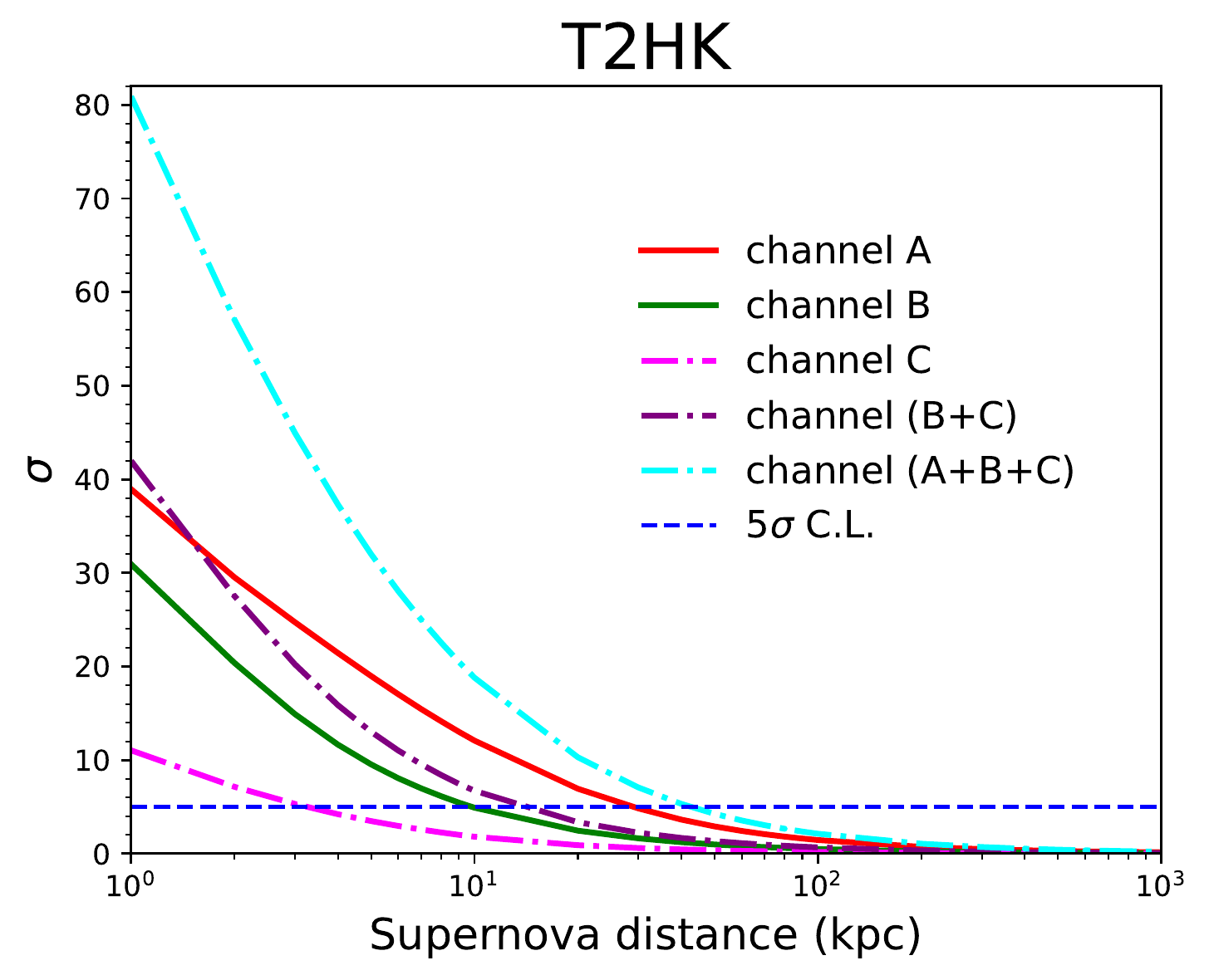}\label{channel-distance-t2hk}}
     \caption{Neutrino mass ordering sensitivity as a function of supernova distance (in kpc). Panel \textbf{(\ref{channel-distance-dune})} is for DUNE and panel \textbf{(\ref{channel-distance-t2hk})} is for T2HK. In each panel, red, green, magenta, purple and cyan curves represent channel A, channel B, channel C, channel (B+C), channel (A+B+C) respectively. Blue dotted line shows the benchmark $5~\sigma$ sensitivity.}
     \label{channel-distance}
\end{figure}

In this subsection, we will present the neutrino mass ordering sensitivity as a function of supernova distance. We will do this for the different detection channels of the supernova neutrinos in T2HK and DUNE. Fig. \ref{channel-distance} shows the variation of the neutrino mass ordering sensitivity as a function of supernova distance (in kpc) for different channels. For both the panels, red, green, magenta, purple and cyan curves are for channel A, channel B, channel C, channel (B+C) and channel (A+B+C) respectively. Blue dotted line denotes the benchmark value of $5 ~\sigma$ C.L. Panel \ref{channel-distance-dune} and panel \ref{channel-distance-t2hk} demonstrate the sensitivity for DUNE and T2HK respectively. We consider a 5\% systematic error corresponding to normalization and energy calibration errors in the event rates in generating these figures. 

From both the panels and all the curves, we see that as the supernova distance increases, the sensitivity decreases. This is because event number of supernova neutrinos depend on supernova distance by inverse square law. Therefore, when supernova distance increases, number of events decreases and as a result, the overall sensitivity decreases. The neutrino mass ordering sensitivity is completely lost if the supernova distance is around 200 kpc. We also see that the overall sensitivity is highest when we add all the three channels (A+B+C) for both the experiments. In general, the sensitivity in T2HK is higher than DUNE. For DUNE, among the three channels, the major contribution towards the determination of neutrino mass ordering comes from the channel A. The sensitivity of channel B is lower than the sensitivity from channel A and the sensitivity of channel C is less than channel B. The sensitivity of channel (B+C) is lower than the sensitivity of channel A. For T2HK, in panel~\ref{channel-distance-t2hk}, we see that, among the three channels, the sensitivity of channel A is greater than the sensitivity of channel B and channel C, similar to DUNE. The sensitivity of channel (B+C) is higher than the sensitivity of channel A, until the supernova distance around 1.8 kpc and after that, the sensitivity of channel A is higher than channel (B+C). 

As a summary from this section, we can understand that by analyzing the data from the supernova neutrinos, one can determine the neutrino mass ordering with at least $5 ~\sigma$ C.L. if the supernova distance is 42.7 kpc for T2HK and 15.2 kpc for DUNE, when the data from all the three channels are combined.

\subsection{Effect of systematics}
\label{effect:sys}

Let us now discuss the effect of systematic uncertainties on the measurement of neutrino mass ordering from the supernova neutrinos. The source of systematic errors arise mainly from the uncertainties in neutrino flux, cross-section measurement, direction of the incoming neutrinos etc. Presence of systematic errors can cause a significant difference in the overall sensitivity. When the supernova distance is small, number of events become very large. As systematic errors directly proportional to the number of events, the effect of systematic uncertainties on the neutrino mass ordering sensitivity can be very large at smaller supernova distance. For large supernova distance, the number of events decreases and therefore, the effect of systematics on the neutrino mass ordering sensitivity decreases simultaneously. Here in our calculation, we have considered two types of systematic uncertainties, one is normalisation error and other is energy calibration error. In this section, we will see the effect of these two systematic errors on the mass ordering sensitivity. For incorporating the above two types of systematic errors, we have used eq. \ref{error} and eq. \ref{chi-sys}.

\begin{figure}[htbp]
      \centering
     \subfloat[] {\includegraphics[height=60mm, width=80mm]{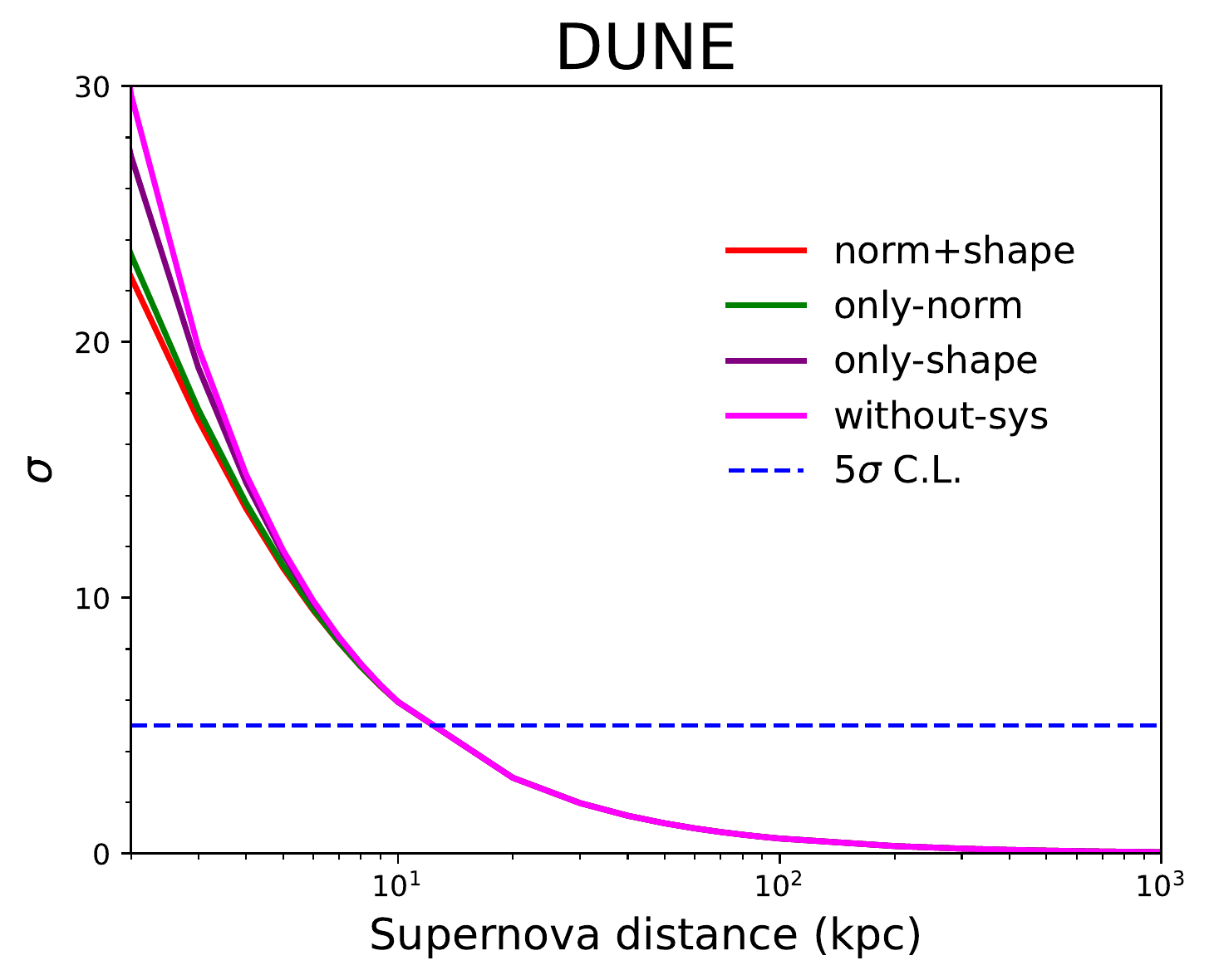}\label{dune-sys}}
      \subfloat[]{\includegraphics[height=60mm, width=80mm]{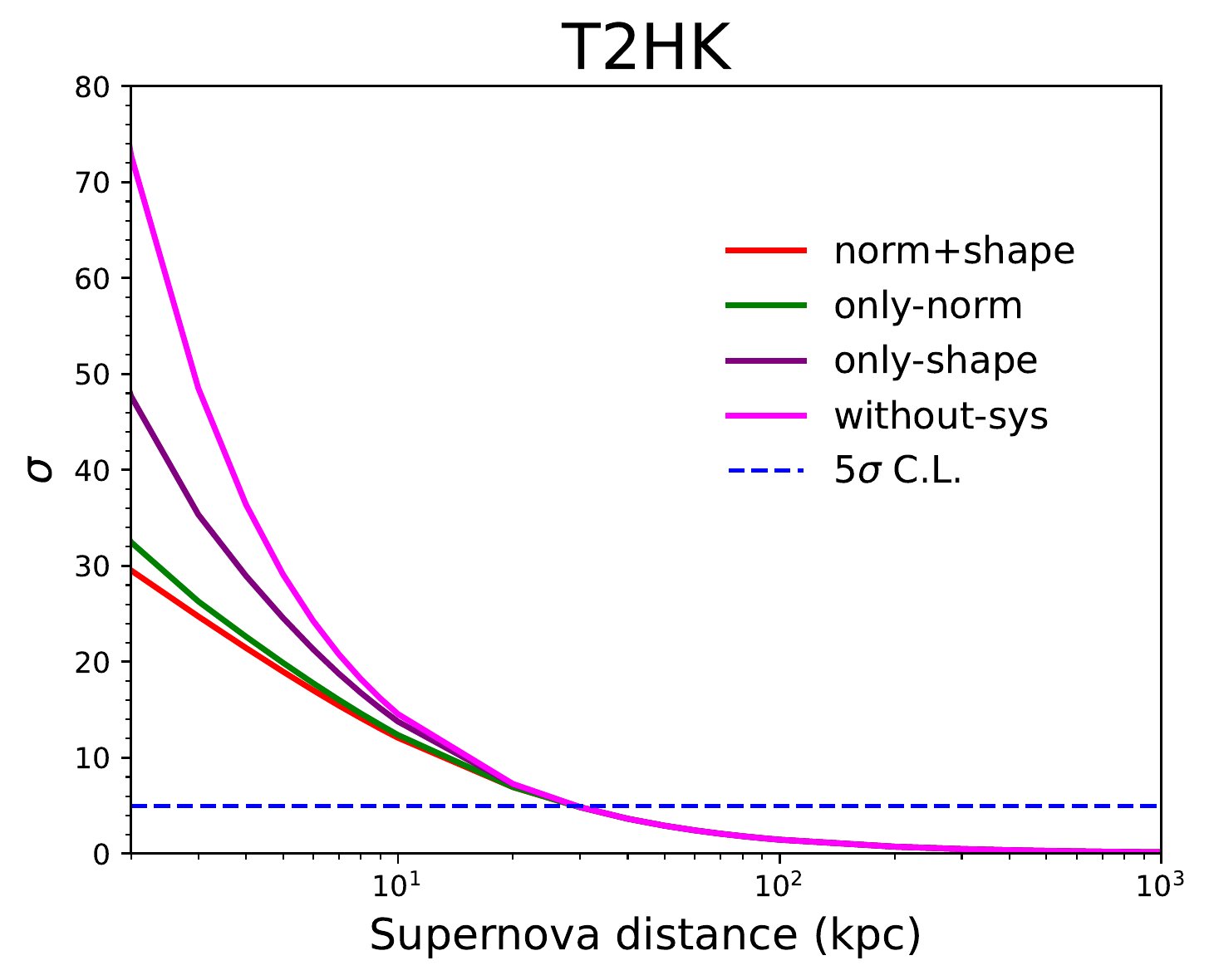}\label{t2hk-sys}}
      \caption{Left panel: mass ordering sensitivity with respect to supernova distance (in kpc) for DUNE. Right panel: mass ordering sensitivity with respect to supernova distance (in kpc) for T2HK. For each panel,``norm" and ``shape" refer to ``normalisation error" and ``energy calibration error" respectively. In each panel, color code is given in the legend. 
      }
      \label{sys-with-all-norm}
      \end{figure}

Fig.~\ref{sys-with-all-norm} shows the effect of systematics on sensitivity for DUNE and T2HK experiments. Left panel of Fig.~\ref{sys-with-all-norm} is for DUNE experiment and right panel is for T2HK. We plot the figure only for channel A as it is the main channel for both  the experiments. In each panel, red curve shows the sensitivity with respect to supernova distance in presence of normalisation as well as energy calibration errors. Green curve shows the result when we  consider only normalisation error, and finally purple curve represents the sensitivity variation in presence of energy calibration error only. In both the panels, magenta curve depicts the condition when we do not consider any type of systematic errors. From Fig.~\ref{sys-with-all-norm}, we can see that, when supernova distance is small,
the effect of normalisation error is more than energy calibration error for both the panels. For large supernova distance, the effect of different systematics become negligible. 

From panel \ref{dune-sys}, for DUNE, we can see that, when the supernova distance is about 9 kpc, the curves with different systematic errors and without systematic error are exactly overlapping. This is because for DUNE, the number of events are not very large (cf. Tab~\ref{event-table}). When the distance is smaller (less than 9 kpc), the effect of normalisation error is more than energy calibration error and when we consider both systematic errors simultaneously, the overall sensitivity decreases significantly.

For T2HK, panel \ref{t2hk-sys}
 shows  similar effect of systematics on sensitivity as that of DUNE. For T2HK, we can see that, after 20 kpc, the effect of systematics are negligible. At smaller distance (less than 20 kpc), the sensitivity due to normalisation and energy calibration errors is lower than other two conditions, i.e., in presence of normalisation error only and in presence of energy calibration error only. From panel \ref{t2hk-sys}, we can see at supernova distance 10 kpc, the curve without systematics has the sensitivity $14.4 \sigma$ whereas its value decreases to $12 \sigma$ when we consider both types of systematic errors. At the same supernova distance, the sensitivity value is $13.8 \sigma$ when we have taken only energy calibration error as systematic uncertainty, and its value becomes $12.4 \sigma$ when only normalisation error is taken into consideration.  
 
    \begin{figure}
        \centering {\includegraphics[height=60mm, width=80mm]{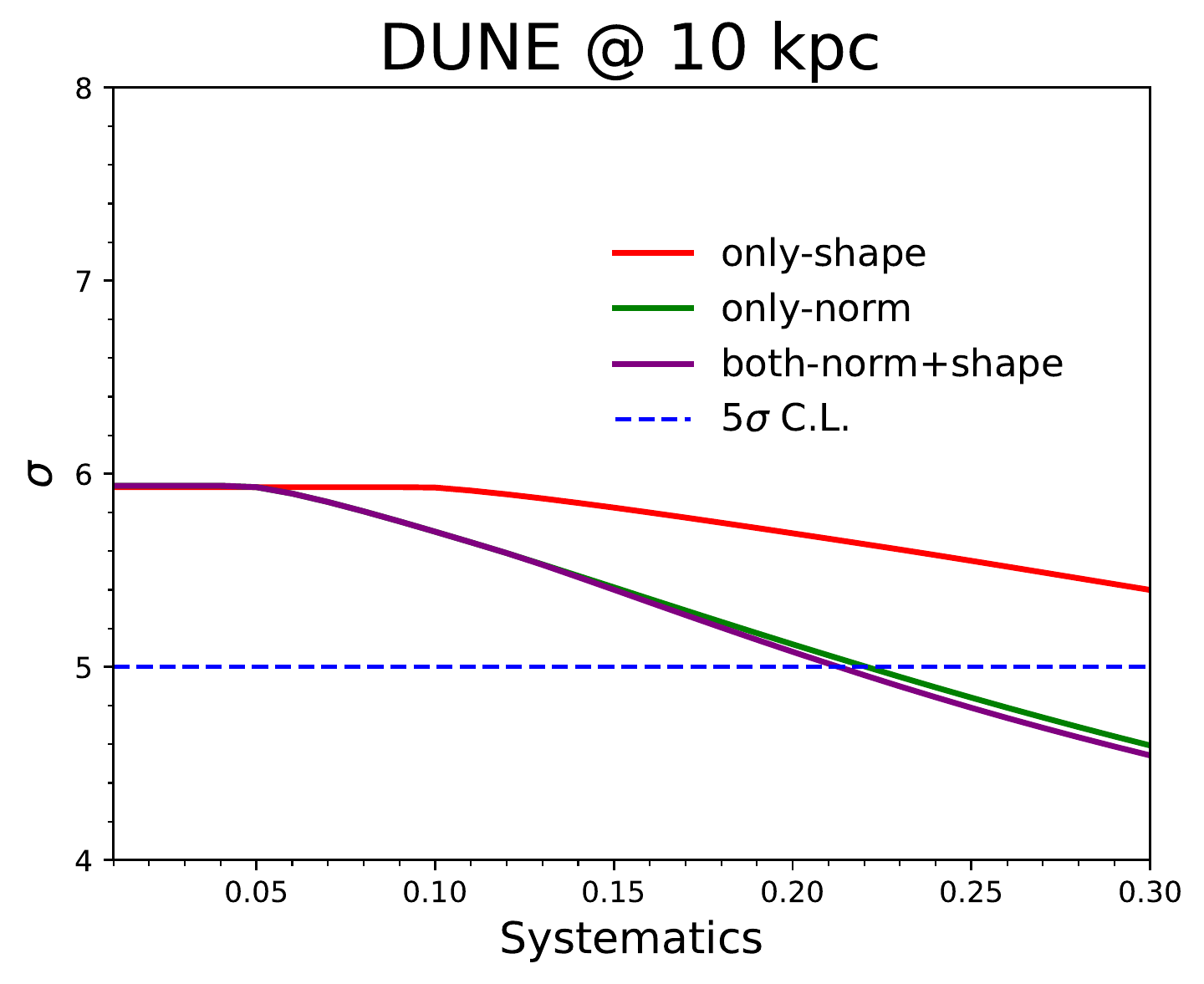}}
      \includegraphics[height=60mm, width=80mm]{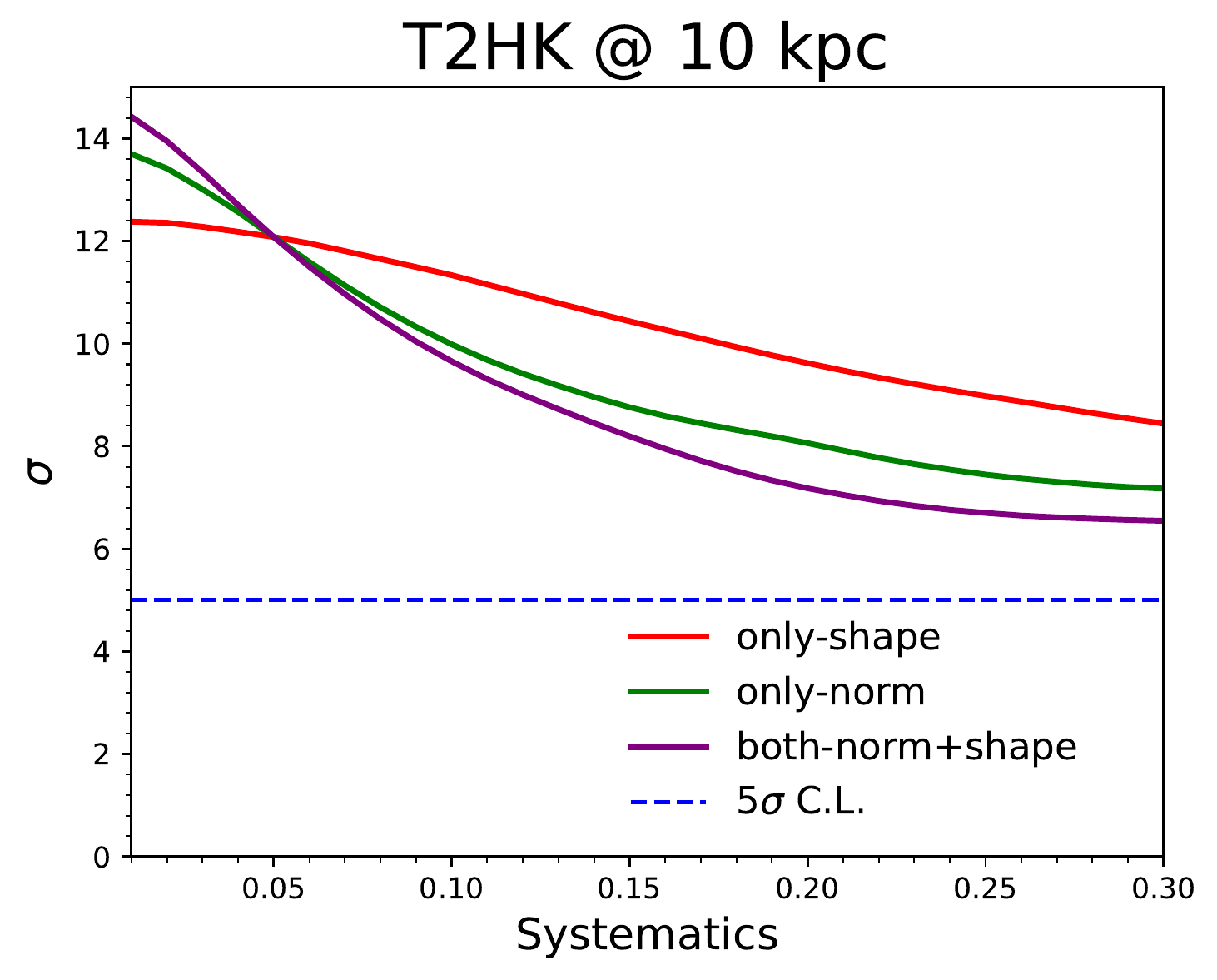}
      \caption{Mass ordering sensitivity at supernova distance 10 kpc as a function systematic error. We have taken three conditions of systematic errors. For each panel, color code is given in legend. Each panel has ``norm" and ``shape" which refer to ``normalisation error" and ``energy calibration error" respectively.}
      \label{sys-1}
 \end{figure}

To understand the variation of the sensitivity with different systematic errors, Fig. \ref{sys-1} has been introduced. This panel is generated for the supernova distance of 10 kpc and for the channel A. Left panel is for DUNE and right panel is for T2HK experiment. In this figure, red curve represents the change of sensitivity when only energy calibration error changes from 0.01 ($1 \%$) to 0.3 ($30 \%$)  and the normalisation error is fixed at 5\%. Similarly, green curve depicts the variation of sensitivity when only normalisation error changes from $1 \%$ to $30 \%$  and the energy calibration error is fixed at 5\%. Purple curve shows the effect of systematics when both normalisation as well energy calibration errors are taken into consideration in the variation from $1\%$ to $30 \%$. 
 From the panels we see that sensitivity decreases as the systematic error increases. We also realise that the effect of normalization error is more as compared to the effect of energy calibration error as the purple curve and the green curve are very close to each other. For DUNE, the systematic errors only play a role if the systematic errors are less than 5\% at 10 kpc. For T2HK, we see all the three curves meet at the y-axis value of 5\%. This is because at this point, all the three curves has the same values systematic errors i.e., 5\% normalisation and 5\% energy calibration. From Fig.~ \ref{sys-1}, we can see that for DUNE (T2HK) the sensitivity falls from $5.9 \sigma$ ($14.4 \sigma$) to $4.5 \sigma$ ($6.5\sigma$) as systematics increases from $1 \%$ to $30 \%$ if one considers both normalisation and energy calibration errors as systematic uncertainties.

\subsection{Effect of smearing }
\label{effect:smear}

\begin{figure}[htbp]
     \centering
     \subfloat[]{\includegraphics[height=60mm, width=80mm]{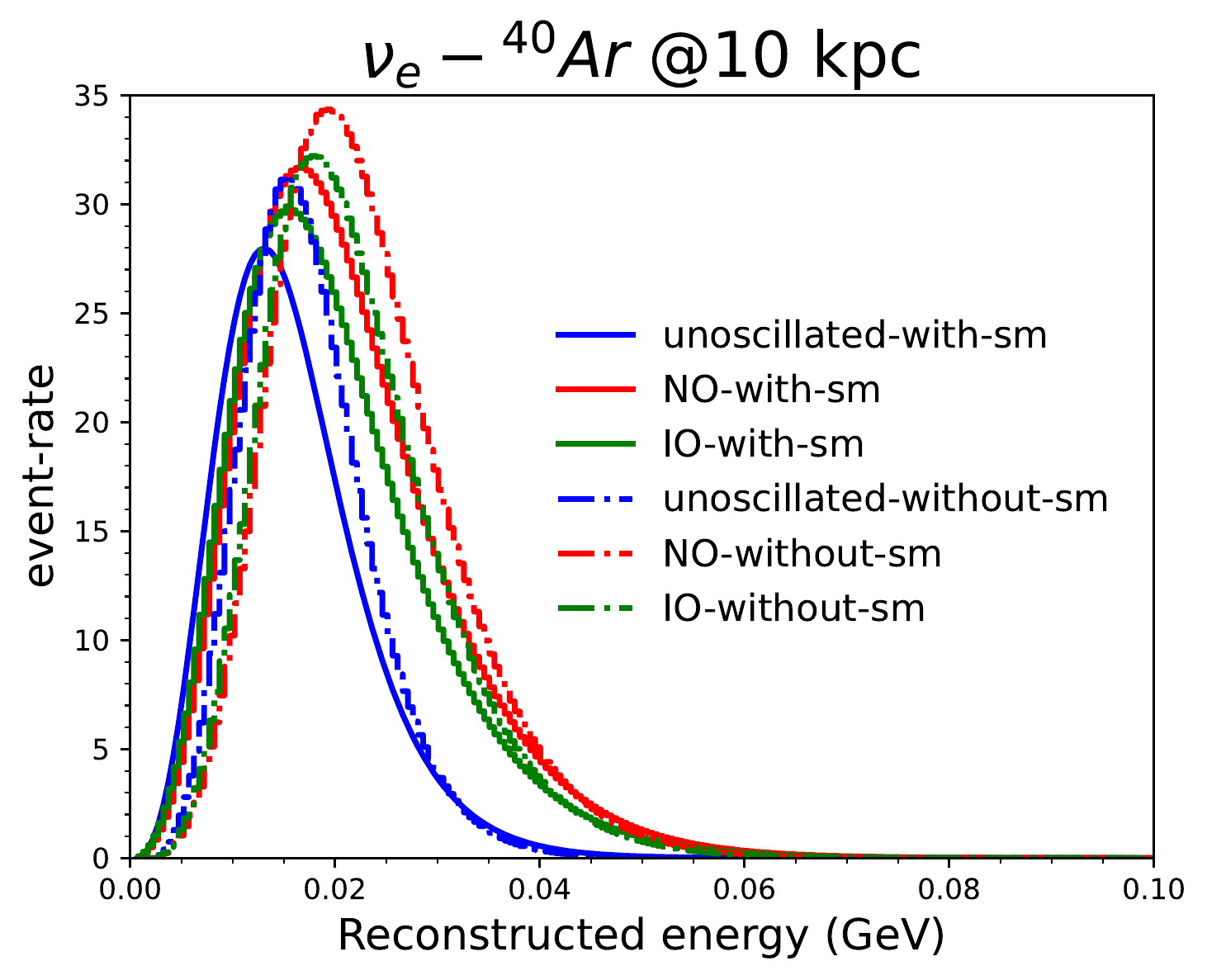}\label{event-smear-dune}}
     \subfloat[]{\includegraphics[height=60mm, width=80mm]{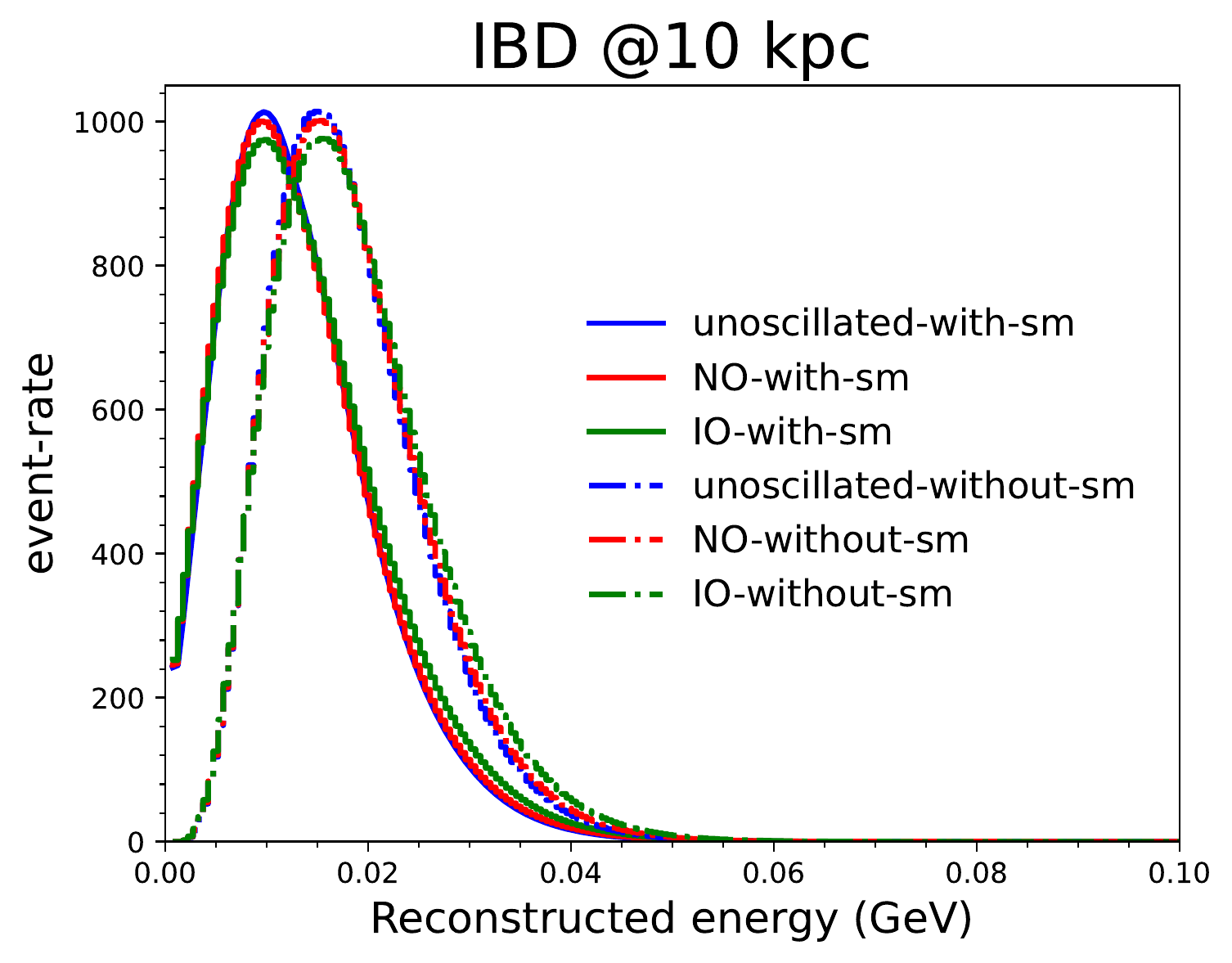}\label{event-smear-t2hk}}
     \caption{\textbf{(\ref{event-smear-dune})} depicts the event rates vs neutrino energy (in GeV) with and without the energy resolution effect for DUNE. \textbf{(\ref{event-smear-t2hk})} represents the same as \textbf{(\ref{event-smear-dune})}  for T2HK. Both the plots are generated for  the supernova distance of 10 kpc.
     For each panel, ``NO-with(without)-sm" refers to ``normal ordering-with (without)- smearing", similarly ``IO-with(without)-sm" defines to ``inverted ordering-with (without)- smearing".
     In each plot, color code is written in legend. 
     }
     \label{event-smear}
\end{figure}

In this subsection, we will discuss the effect of energy resolution in DUNE and T2HK. In DUNE and T2HK, the energy of the neutrinos will be reconstructed by measuring the energy and momentum of the outgoing leptons. In our analysis, we incorporate this effect by the inclusion of energy resolution. Because of this energy resolution, the neutrino events will be smeared around its true energy causing a loss of information. Therefore, in the presence of energy smearing, the sensitivity expected to become worse as compared to the sensitivity without energy smearing. To show this in Fig. \ref{event-smear} we plot the event spectrum of the supernova neutrinos with and without energy smearing. Panel~\ref{event-smear-dune} (Panel~\ref{event-smear-t2hk}) shows the event numbers as a function of neutrino energy for a 10 kpc supernova distance in DUNE (T2HK) considering the main channel i.e., channel A. In each panel, blue solid (blue dot-dashed), red solid (red dot-dashed) and green solid (green dot-dashed) curves are the event rates with (without) the presence of smearing. From this figure we note that, in the presence of energy smearing, the whole event spectrum shifts towards left i.e., in the lower energy region causing a change in the shape of the spectrum. For DUNE, we notice a change in the height of the spectrum. The spectrum corresponding to no energy smearing is higher as compared to the spectrum with smearing. However, the energy resolution factor, does not change the normalization of the event spectrum in T2HK.

\begin{figure}[htbp]
     \centering
     {\includegraphics[height=70mm, width=90mm]{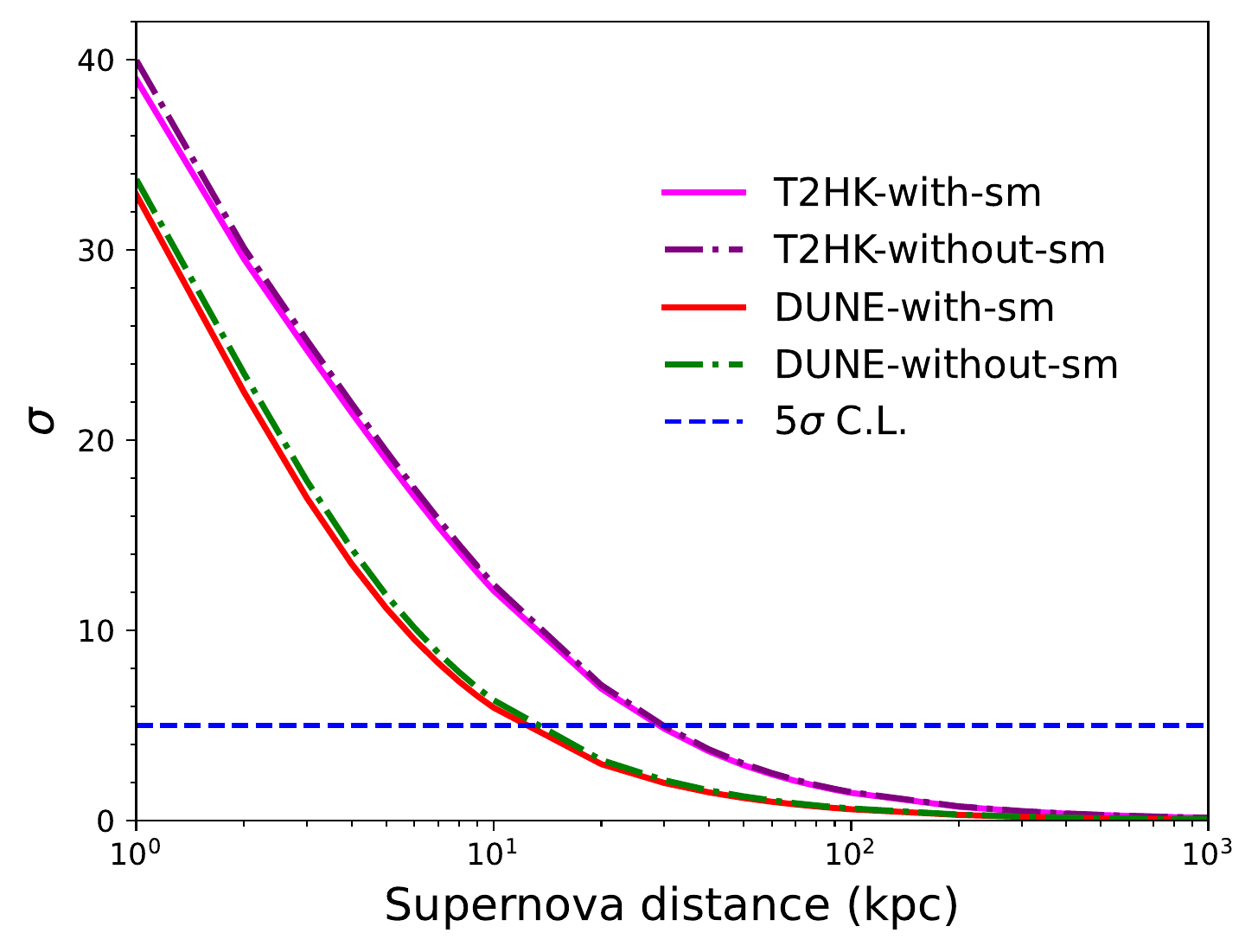}}
     \caption{Mass hierarchy sensitivity as a function of supernova distance (in kpc) for  DUNE and T2HK. In each panel, color code is given in legend. For each panel,
``sm” refers the term ``smearing".
     }
     \label{smear}
\end{figure}

Next, let us see how the hierarchy sensitivity gets modified because of the energy smearing. Fig. \ref{smear} shows mass ordering sensitivity with respect to supernova distance in absence and presence of smearing. In this figure, red solid (green dot-dashed) and magenta solid (purple dot-dashed) curves show the results for DUNE and T2HK in presence (absence) of energy smearing respectively. This figure is generated for the channel A of both DUNE and T2HK, and considering a $5\%$ systematics of both normalisation and energy calibration errors. As expected, from the figure, we can see that in presence of smearing, the sensitivity is less compared to the sensitivity in absence of smearing. This is true for all the values of supernova distance and for both the experiments. At a supernova distance of 10 kpc, the deterioration in the sensitivity due to energy smearing is around $0.4~ \sigma$ for DUNE and $0.3 ~\sigma$ for T2HK.

\vspace{-0.2cm}
\section{Summary and conclusion}
\label{summary}

In this work, we have studied the neutrino mass ordering sensitivity of the T2HK and DUNE detectors by analyzing the expected event rates of a future supernova explosion. First, we showed the time integrated neutrino flux spectrum of supernova corresponding to $\nu_e$, $\bar{\nu}_e$ and $\nu_x$ assuming the Garching model of the supernova. Then we calculated the expected event rates corresponding to different detection channels in DUNE and T2HK detectors. In DUNE, the supernova neutrinos will be mainly detected by the following three processes: $\nu_e-^{40}\rm Ar $ (channel A), $\bar{\nu}_e-^{40}\rm Ar $ (channel B) and $\nu_e-e$ (channel C). Whereas, in T2HK, the leading detection channels for the supernova neutrinos will be: IBD (channel A), $\bar{\nu}_e-^{16}O$ (channel B) and $\nu_e-e$ (channel C). As the conversion probability of the neutrinos inside the supernova is different for normal and inverted orderings, it is possible to determine the true ordering of the neutrino masses by analyzing those event rates which are different for different ordering of the neutrinos. In our analysis, we have considered only MSW transitions. In order to estimate the neutrino mass ordering sensitivity from the event rates, we have defined a Poissonian $\chi^2$ formula and included  5\% systematic error corresponding to both normalisation as well as energy calibration error of the theoretical events. We then plotted this $\chi^2$ as a function of the supernova distance for the different detection channels and their combinations. In this context, we have also studied the effect of systematic error and the effect of energy smearing on the measurement of mass ordering sensitivity from the supernova neutrinos.

\begin{table}[htpb]
%\hspace{-1.2cm}  
\begin{tabular}{|l|l|l|l|l|l|l|l|l|}
    \hline
    \rowcolor{citrine!10}
    \multirow{2}{*}{Setup} &
      \multicolumn{1}{c|}{SD} &
      \multicolumn{1}{c|}{SD} &
      \multicolumn{1}{c|}{SD}  &
      \multicolumn{1}{c|}{SD}  &
      \multicolumn{1}{c|}{SD} \\
    &Channel A & Channel B &Channel C &Channel (B+C) & Channel (A+B+C) \\
    \hline
    \rowcolor{emerald!20}
DUNE & ~12.4 kpc  & ~~1.9 kpc &~~~ ~~-   & ~~~~~2.5 kpc   & ~~~~~15.2 kpc   \\
    \hline
    \rowcolor{fulvous!30}
    T2HK & ~29.1 kpc & ~~9.8 kpc & ~~3.2 kpc  & ~~~~14.2 kpc  & ~~~~~42.7 kpc      \\
    \hline
  \end{tabular}
  \caption{Supernova distances for which a $5 \sigma$ neutrino mass ordering sensitivity can be achieved. For the table, we have considered the systematic errors for both normalisation and energy calibration.}
  \label{table_sum}
\end{table}

We summarize our main results in table~\ref{table_sum}. This table shows the supernova distance for which a neutrino mass ordering sensitivity of $5 ~\sigma$ can be achieved. This is shown for both DUNE and T2HK corresponding to individual detection channels and their combinations. For each channel, the supernova distance is given for the case when both systematic errors are taken into consideration. Our results show that it is possible to determine neutrino mass ordering by analyzing the oscillated $\nu_e$ and $\bar{\nu}_e$ events from the supernova. Among the three channels, the dominant contributions come from channel A for both DUNE and T2HK. However, the best sensitivity comes when all the three channels are combined together. We have shown that the systematic errors affect more in the lower supernova distances where the sensitivity is dominated by statistics. The effect of systematics is more in T2HK as compared to DUNE. Between the normalisation error and energy calibration error, the deterioration of the sensitivity is mostly dominated by the normalisation error.

Between DUNE and T2HK, the sensitivity of T2HK is several times higher because of the large detector volume of the T2HK detector. Finally we also show that the sensitivity of DUNE and T2HK will be deteriorated to some extent because of the energy smearing which will arise due to the energy reconstruction of the supernova neutrinos.

In summary, we would like to emphasize that, if a supernova explosion happens during the running time of DUNE and T2HK, these detectors provide an excellent opportunity to determine the true nature of neutrino mass ordering. For the most optimistic case i.e., when data is combined from the all the three channels, we expect that  the neutrino mass ordering can be  determined at $5 ~\sigma$ C.L., if the supernova explosion occurs at a distance of 42.7 kpc for T2HK and 15.2 kpc for DUNE. This is true if we assume a 5\% systematic error in our analysis.

\section*{Acknowledgements}

PP wants to acknowledge Prime Minister’s Research Fellows (PMRF) scheme
for financial support. RM would like to acknowledge University of Hyderabad IoE project grant no: RC1-20-012. We gratefully acknowledge the use of CMSD HPC facility of University of Hyderabad to carry out the computational works. This work has been in part funded by Ministry of Science and Education of Republic of Croatia grant No. KK.01.1.1.01.0001. We also acknowledge Dinesh Kumar Singha, Samiron Roy and Riya Gaba for useful discussions.

\bibliography{supernova}
\end{document}